\documentclass[12pt,preprint]{aastex} 

\received{\today} 
\begin{document} 

\title{{\it Chandra} Observations of the Eagle Nebula.
I. Embedded Young Stellar Objects near the Pillars of Creation} 

\author{Jeffrey L. Linsky\altaffilmark{1},
Marc Gagn\'e\altaffilmark{2}, Anna Mytyk\altaffilmark{2},
Mark McCaughren\altaffilmark{3} \& Morten Andersen\altaffilmark{4}}
\email{\small jlinsky@jila.colorado.edu, mgagne@wcupa.edu}

\altaffiltext{1}{JILA, University of Colorado and National Institute of 
Standards and Technology, Boulder, CO 
80309-0440} 
\altaffiltext{2}{Department of Geology and Astronomy, West Chester University, 
West Chester, PA 19383} 
\altaffiltext{3}{School of Physics, University of Exeter, Stocker Road, 
Exeter, UK EX4 4QL}
\altaffiltext{4}{Steward Observatory, University of Arizona, 933 N. Cherry 
Ave., Tucson, AZ 85721}

\begin{abstract} 

We present and analyze the first high-resolution 
X-ray images ever obtained of the Eagle Nebula
star-forming region. On 2001 July 30 the {\it Chandra} X-ray Observatory
obtained a 78 ks image of the Eagle Nebula (M~16) that includes
the core of the young galactic cluster NGC~6611 and the dark
columns of dust and cold molecular gas in M~16
known as the ``Pillars of Creation''.  We find a total of 1101
X-ray sources in the $17\arcmin\times17\arcmin$ ACIS-I field of view.
Most of the X-ray sources are low mass pre-main-sequence or
high-mass main-sequence stars in this young cluster.
A handful of hard X-ray sources in the
pillars are spatially coincident with deeply embedded young stellar 
objects seen in high-resolution near-infrared images recently obtained
with the VLT \citep{MA2002}.
In this paper, we focus on the 40 X-ray sources in and around
Pillars 1--4 at the heart of the Eagle Nebula.
None of the X-ray sources are associated with the
evaporating gaseous globules (EGGs) first observed by \citet{Hester1996}
in {\it HST} WFPC2 images of M~16, implying that either the EGGs do not contain
protostars or that the protostars have not yet become X-ray active.
Eight X-ray counts are coincident with the Herbig-Haro object HH216,
implying $\log L_{\rm X}\approx 30.0$.

\end{abstract} 


\section{Introduction}

\citet{Hester2004} presented evidence suggesting that
the Sun was born in a massive star-forming region.
For example, the discovery of radioactive $^{60}$Ni, a decay product of 
radioactive $^{60}$Fe, in two chondritic meteorites \citep{Tachibana2003}
implies that $^{60}$Fe was present when the Ca-Al-rich inclusions (CAIs)
were formed 4.5 Gyr ago. Because $^{60}$Fe cannot be formed efficiently
by particle irradiation, its presence requires input from a nearby
supernova, which prior to the explosion was a massive O star.
Because the half-life of $^{60}$Fe is only 1.49 Myr,
the CAIs must have formed within a few million years of a nearby
supernova explosion. While the star-forming regions within 200~pc
of the Sun are generally classified as low-mass star-forming regions
(Taurus-Auriga, Ophichus, Upper Scorpius), it is not clear whether
most stars in the Galaxy are formed in such small clouds and loose associations
rather than in high-mass star-forming regions like the Orion Nebula and the
Eagle Nebula (M16), the subject of this and subsequent papers of the series.

Clusters of young stellar objects (YSOs) emerging from their 
molecular clouds are the laboratories of
choice for addressing fundamental questions of star formation. 
Such clusters include YSOs with
different masses, ages, and distances from hot O stars. 
These clusters often contain YSOs in diverse environments, including 
(a) completely embedded very young Class~0 objects visible only at 
millimeter and sub-millimeter wavelengths, (b) partially embedded Class~I
sources visible at infrared and X-ray wavelengths, (c) Class~II T Tauri 
stars with circumstellar disks and their more massive cousins,
the Herbig Ae/Be stars, and (d) Class~III naked or weak-lined T Tauri stars 
that have completely emerged from their nascent cloud environments.
Reviews of young stellar clusters and their molecular cloud
environments include 
the physical conditions in star-forming regions \citep{Evans1999}, 
high energy processes in young stellar objects \citep{Feigelson1999}. 
the important case study of the Orion Nebula \citep{Odell2001}, 
dusty circumstellar disks \citep{Zuckerman2001}, 
interaction of stellar jets with the molecular gas cloud 
\citep{Reipurth2001},
clusters embedded in molecular clouds \citep{Lada2003}, and
young stars near the Sun \citep{Zuckerman2004}.

Both low-mass and high-mass star-forming regions
can be studied at a variety of wavelengths from radio waves to
X-rays. Observations at radio wavelengths, in H$\alpha$ and other
emission lines are useful for studying the gas in \ion{H}{2} regions ionized
by nearby O stars. Observations in the near- to far-infrared are useful for
identifying deeply embedded YSOs and their cold disks. X-ray imaging of
clusters is valuable for identifying cluster membership since YSOs have
very high X-ray luminosities, as large as
$\log L_{\rm X}/L_{\rm bol}\approx 10^{-3}$, and they can be detected through
column densities as large as  N(H~I) $\approx 10^{23}$~cm$^{-2}$. Typically, 
most of the X-ray sources in fields centered on a cluster are cluster members 
rather than foreground or background stars or AGN.

Beginning with the {\it Einstein} X-ray Observatory and continuing with
{\it ROSAT}, {\it ASCA}, XMM-{\it Newton}, and {\it Chandra}, 
X-ray imaging of star-forming
clusters has demonstrated
that YSOs, including class I protostars and the older
class~II (T-Tauri) and class~III (naked T Tauri) stars,
are bright X-ray sources with luminosities
$L_{\rm X}/L_{\rm bol} \gtrsim 10^{-4}$ outside of flares and
considerably larger during flares.
As a result of its unprecedented high angular resolution, low background
and high throughput, the {\it Chandra} X-ray Observatory is the
instrument of choice for studying crowded star clusters. 
Recent examples of clusters studied with {\it Chandra} include
$\rho$~Oph \citep{Gagne2004}, Chamaeleon I North Cloud \citep{Feigelson2004},
NGC~2264 \citep{Ramirez2004}, and R136 \citep{Townsley2006},

The most comprehensive study of a young massive star-forming region in X-rays 
is the {\it Chandra} Orion Ultradeep Project (COUP), an 838-ks exposure
of the Orion Nebula Cluster (ONC) with ACIS-I. A total of
1315 X-ray sources were detected from confirmed Cluster members 
with masses extending from brown
dwarfs to the central 40~M$_{\odot}$ O star and with X-ray luminosities in 
the range of
$\log L_{\rm X} =$ 27.3 -- 33.3 ergs s$^{-1}$.  The results of this study
are described in a series of 13 papers in the October 2005 issue of the 
{\em Astrophysical Journal Supplement Series} beginning with 
\citet{Getman2005a} and \citet{Getman2005b}. 
Two perhaps surprising results of this
study are (i) the presence or absence of a circumstellar disk appears to have
no discernable effect on the X-ray emission and (ii) the plasma temperatures 
of T Tauri stars are often very high ($T \geq 100$ MK), even when no flares are
detected.

In a new series of papers, we present and analyze observations of M~16,
another massive star-forming region also known as the ``Eagle Nebula'',
which is an \ion{H}{2} region
excited by the  hot stars in the cluster NGC~6611.  
The nebula is located at a distance of $2.0\pm 0.1$ kpc,
based on spectroscopic paralaxes \citep{Hillenbrand1993} or $2.14\pm 0.1$ kpc, 
based on fitting stars in the upper part of the color-magnitude diagram to 
the zero age main sequence \citep{Belikov1999}. Star formation in NGC~6611
likely proceeded over a considerable time period as \citet{Hillenbrand1993} 
estimate an age of about 2~Myr for the the massive star population 
but say that the lower mass PMS stars range in age from 0.25~Myr to greater 
than 1Myr. \citet{Hillenbrand1993} and \citet{Belikov2000} also find  
that the oldest stars in the cluster have ages of about 6~Myr. 
The very youngest stars in M~16 are those embedded in or at the edge of 
the ``Pillars of Creation.''

\subsection{Morphology of the ``Pillars of Creation''}

The 1995 WFPC2 image of
the ``Pillars of Creation" \citep{Hester1996}
is arguably the most famous image obtained by the
{\it Hubble Space Telescope (HST)}.
The morphology of Pillars 1, 2 and 3 is presented in detail through
the WFPC2 images obtained by \citet{Hester1996} in H$\alpha$,
[S II] $\lambda 6717, 6731$ + continuum, and [O III] $\lambda$5077.
The WFPC2 image of Pillars 1-3 (from left to right) 
is reproduced in Figure~1 (courtesy NASA and the STScI). Figure~2 shows 
the same field but obtained with the near-infrared $J_{\rm s}$, $H$, 
and $K_{\rm s}$ bands by \citet{MA2002}. In both figures the detected 
{\it Chandra} sources are shown as circles.
Pillar 4 is located $3\arcmin$ southeast of Pillars 1-3 and has not been
imaged with {\it HST}. The location of Pillar 4 is shown in Figure~3.

The ongoing destruction of the pillar molecular clouds is 
evident from $^{12}$CO 
line profiles which show velocity gradients of 20~km~s$^{-1}$~pc$^{-1}$ 
pointing directly back to the
O stars of NGC 6611 (NW of the pillars) and by the agreement of the 
pillar morphology with hydrodynamic models \citep{Pound1998, Pound2005}.
The H II region is ionized primarily by one massive star,
NGC 6611\footnote{NGC 6611 source numbers 
in the catalogs of \citet{Walker1961}, \citet{Kamp1974}, \citet{Tucholke1986}, 
\citet{Hillenbrand1993}, and \citet{Belikov1999}. \citet{Evans2005} gives
slightly different spectral types for these stars.}
205 = HD~168076 O5~V((f*)), but also by three other O stars: 
175 = BD$-13\arcdeg 4923$ O5.5~V((f)),
197 = HD~168075 O7~V((f)), and
246 = BD$-13\arcdeg 4927$ O7II(f). Figure~3 shows the relationship of
Pillars 1--3 to the O stars near the center of NGC~6611.

Near-infrared \citep{Sugitani2002} and millimeter \citep{Fukuda2002}
observations show that the pillars have head-tail morphologies 
(cf. Pound 1998) in which the
most opaque portions are located in the heads facing toward the massive O star
(HD~168076) and the less opaque tails face away from the O star. 
This structure and the
observed blue- and red-shifted millimeter emission are consistent with the
picture of the pillar stucture being shaped by the photoevaporative flow
from the O stars. 

The $JHK_{\rm s}$ survey of \citet{Sugitani2002} of a $2\arcmin\times3\arcmin$
region centered on Pillars 1--3 
represents the most comprehensive attempt to date
to identify the low-mass stellar population of M~16. The resulting
near-infrared color-color diagram allowed them to identify three populations 
of stars (see \S3.1):
moderately reddened stars between the giant and dwarf redenning vectors 
(``F'' sources),
reddened T Tauri stars with moderate $H-K$ excesses (``T'' sources) and
protostars with $\Delta(H-K) > 1.0$ (``P sources").
The P and T sources are generally seen near the tips and edges of Pillars 1-3.
The most famous of these sources are P1=M16ES-1=YSO~1 in the tip of
Pillar~1 and T1=M16ES-2=YSO~2 in tip of Pillar~2.

Observations with the Near Infrared Camera and Multi-Object Spectrometer
(NICMOS) on {\it HST} led \citet{TSH2002} to identify two active regions of
star formation in M~16 located at the tips of Pillars 1 and 2.
Submillimeter \citep{White1999} and millimeter \citep{Fukuda2002} images
support this conclusion. The youngest YSOs are located near the pillar tips
where the interaction between the ionizing radiation from the O stars is most
intensely impacting the molecular cloud and fashioning its structure 
\citep{Sugitani2002}.
\citet{MA2002} estimate that M16ES-1 is a $10M_\odot$ YSO with
$A_V\approx27$. M16ES-2 is most likely a 1~Myr old, $2-5M_{\odot}$
YSO with $A_V\approx15$. These extinction estimates will be used
to derive $L_{\rm X}$ and $L_{\rm bol}$ in \S3.3 and \S3.4.
It is likely that the strong mass flow and radiation from the O stars has also
produced shocks in the pillars that have triggered the formation of these YSOs
\citep{Elmegreen1992,TSH2002}. These stars have presumably just emerged from
the pillars as a result of the photoevaporative flow.

Although the near-infrared color-color diagram can be used to identify
some Class I and Class II YSOs, near-infrared photometry alone does
not easily distinguish non-members from cluster members without large
K-band excesses. As a result, most of the low-mass stellar population
has yet to be catalogued.

\subsection{Evaporating Gaseous Globules (EGGs) in the Eagle Nebula}

The thin interface between the dense molecular gas in the pillars and the 
surrounding \ion{H}{2} region is
particularly interesting. \citet{Hester1996} argued that the interface is 
produced by photoionization in a photoevaporative flow driven by UV
radiation from the nearby O stars. In effect, the molecular cloud 
is being evaporated, gradually revealing
embedded condensations that may or may not contain YSOs. 
They call attention to the 73
small cometary globules at the evaporating surface or just outside it
(see Figure~1), which
they name ``evaporating gaseous globules,'' or ``EGGs.'' 
\citet{Hester1996} showed that the shapes of the EGGs are as expected from
evaporation by radiation from the O stars with shadowing by the EGG itself. The
geometry of M~16 is particularly favorable for studying these EGGs as they are
viewed from the side, perpendicular to the line of sight radiation from
the O stars. \citet{Hester1996} argue that the EGGs may be physically
the same stuctures as
previously identified as ``proplyds'' \citep{Odell1993} or small nebular
condensations in the Trapezium Cluster in Orion (M~42 = NGC~1976) and other
young clusters. They conclude that the difference in shape between the EGGs in
M~16 and the proplyds in M~42 is due to the different viewing angles -- the
EGGs in M~16 are viewed from the side, whereas the proplyds in M~42 are viewed
parallel to the photoevaporative flow. 
\citet{Hester2004}, on the other hand, argue that EGGs evolve into proplyds.
The ONC proplyds are more easily seen than the M~16 EGGs because the ONC 
proplyds are viewed against a bright background which they shadow 
rather than a dark background for the M~16 EGGs.

\citet{Hester2004} presented a detailed scenario for the formation of 
low-mass stars like the Sun and their associated planetary systems 
from dense molecular clouds illuminated by ionizing UV radiation from young, 
O stars. The pillars in M~16 are likely excellent examples of such 
stellar nurseries developing in this harsh radiation environment.
Their multistage scenario starts with a shock wave that 
compresses the cold gas to create dense protostellar cores, 
followed by an advancing ionization front that photevaporates 
the gas around the dense core producing the EGG phenomenon. Depending on 
circumstances, further evaporation either completely evaporates the core
or evaporates the disk producing proplyds and eventually stars with 
planetary systems. Photoevaporation effectively terminates
the accretion of gas onto the YSO thereby determining its mass when 
it reaches the main sequence.
\citet{Hester2004} estimate that the EGG and proplyd 
stages are short, about $10^4$ years. The Chandra observations provide 
an excellent opportunity to study the high energy aspect of these phenomena.

One test of this scenario for the formation of low-mass stars is to determine
whether the EGGs actually contain YSOs. \citet{MA2002} have done this by 
obtaining very deep images of M~16 centered on the pillars
with the ISAAC near-infrared camera/spectrograph on the VLT (see Figure~2). 
Their 1--2.5$\mu$m
survey reaching to K$_{\rm s} = 20.4$ detected embedded YSOs through up to 30
magnitudes of visual extinction. Of the 73 EGGs identified by 
\citet{Hester1996}, only
11 show definite evidence (and 5 more show tentitive evidence) for an
associated infrared point source in the VLT data, 
indicating the presence of a YSO with mass greater
than 0.02 M$_{\odot}$. Four have masses
in the range 0.3--1 M$_{\odot}$, and the rest are substellar (0.02--0.07
M$_{\odot}$). Millimeter observations also show that not all EGGs contain
emission cores that are identifiable as YSOs \citep{Fukuda2002}. Thus 
many of the EGGs may
not contain YSOs, at least down to the detection limit of the deep
\citet{MA2002} survey, but the presence of EGGs near massive stars is
clear evidence that the harsh radiation environment created by the massive
stars can prematurely halt the formation of YSOs and their associated disks. 
The two embedded sources seen by \citet{Sugitani2002}, \citet{TSH2002} and
\citet{MA2002} are P1=M16ES-1=YSO1 located at the tip of Pillar 1 and 
T1=M16ES-2=YSO2 located at the tip of Pillar 2. Neither of these sources 
are EGGs.

\subsection{Outline of the Eagle Nebula Series of Papers}

Understanding the complex interplay between YSOs, their
nascent molecular clouds, and the ionizing radiation environments 
they encounter is required
for the construction of realistic models describing how stars form and evolve
through their pre-main-sequence phases and eventually arrive on the main
sequence. Many fundamental physical questions must be answered: including the
roles played by the ionizing radiation and shock fronts produced by nearby
massive stars and supernovae,
magnetic fields, angular momentum transfer from the star
to its disk, disk formation and dissipation, mass transfer by accretion, winds,
and jets, the preference for multiple systems, and condensation processes in
the disk leading to planetary formation.

In this first paper of the series, we focus on what we have learned from
X-ray and near-infrared photometry of the YSOs in and around the
``Pillars of Creation''.
In \S2 we describe the {\it Chandra} observations of M~16 and list the 
detected X-ray sources. In \S3.1 and \S3.2, we identify the YSOs near the 
pillars and present our method for deriving the intrinsic X-ray luminosities
of YSOs with corresponding near-infrared photometry. Section \S3.3 explains 
why we think that the massive star at the tip of Pillar 1 is a magnetically 
active YSO likely similar to $\theta^1$~Ori~C (O7~V) in the ONC. Sections
\S3.4, and \S3.5 describe the T Tauri star M16ES-2 and the X-ray emission from
the Herbig-Haro object HH216. In section \S3.6 we show that X-rays are 
not detected from the EGGs and that the nondetection of X-rays from the high 
mass EGGs likely indicates that the YSOs in their cores represent a very 
early stage
of stellar evolution in which magnetic heating processes do not yet 
produce strong X-ray emission.
In subsequent papers, we will study the the \ion{H}{2} regions, 
high and low mass stars in NGC~6611, star formation rates, 
the initial mass function, masers
and cometary clouds in Pillar 5, and other topics.

\section{{\it Chandra} Observation \& Data Reduction}

On 2001 July 30-31, the {\it Chandra} X-ray Observatory observed the Eagle
Nebula (M~16) continuously for 78~ks with a livetime of 77~ks. {\it Chandra} 
collected consecutive
exposures of 3.24~s with the Advanced CCD Imaging Spectrometer (ACIS), using
ACIS chips I0-I3 and S2-S3. The ACIS-S chip data were not used and will not be
discussed in this paper. 
The data reduction, astrometric correction, source detection, and
source matching with the optical proper-motion catalog of \citet{Belikov1999}
and the near-infrared images of \citet{Sugitani2002} and \citet{MA2002}
are discussed here and in more detail in Paper~II.
In this paper we summarize the data reduction process and
focus on the X-ray and infrared sources near Pillars 1--4.

The high-resolution mirror assembly (HRMA) and ACIS-I camera are described in
detail in ``The {\it Chandra} Proposers' Observatory Guide".  The ACIS I3
aim-point was located at (J2000) $\alpha = 18^{\mathrm h}18^{\mathrm m}44\fs0,
\delta = -13^\circ48\arcmin12\arcsec$.  
We applied a standard data reduction procedure using CIAO v2.2.2. 
The ACIS data were obtained in VFAINT mode, allowing us to remove afterglow
events. The event list was filtered to include events with standard ASCA grades
and with photon energies in the 0.5-7.0~keV band, thereby significantly
reducing the particle background above 7~keV.

The {\it Chandra} ACIS-I point source sensitivity degrades off-axis as result 
of vignetting and the larger point-spread function. Source counts were 
extracted using circles (or ellipses) with radii ranging from 3 pixels 
on-axis to 12 pixels at the edge of the CCDs. For example, at the tips of 
Pillars 1--3, the source radii were typically 5 pixels. Background counts were 
subtracted based on blank-sky observations available through the {\it Chandra} 
X-ray Center. Because of the low ACIS-I VF-mode background in the 0.5-7 keV 
band, the background was typically less than 1 count per source (see Table 1).
We rejected sources with 6 or fewer raw counts. 
For the embedded T Tauri star at the tip of Pillar~2 (M16ES-2), we list it 
as detected with only 5 raw counts. The flux at Earth was 
calculated based on the observed photon energies and the computed effective 
area as a function of energy. The corresponding limiting flux depends on the 
source's location and the column density and temperature of the emitting 
plasma. For example, source J181849.2-134938 in Table 1 has a mean
photon energy of only 1.07 keV and a flux at Earth of only 
$1.6\times10^{-16}$~ergs~cm$^{-2}$~s$^{-1}$. In contrast, source 
J181847.9-135058, 
with the same number of counts, has a mean energy of 2.71 keV and a flux at 
Earth of $1.05\times10^{-15}$~ergs~cm$^{-2}$~s$^{-1}$.

The X-ray sources were matched to optical and near-infrared counterparts 
if their 
positions were separated by less than $2\arcsec$, allowing for random position
errors and possible systematic offsets. The RMS offset between {\it Chandra} 
and 2MASS positions was $0.4\arcsec$. In nearly all cases, each {\it Chandra} 
source had a single 2MASS counterpart. In some cases, a single {\it Chandra} 
or 2MASS source was resolved into many sources in the VLT K-band mosaic. 
In nearly all cases the brighter star was closest to the {\it Chandra} 
and 2MASS position.

Fig.~3 compares the $16\farcm9 \times 16\farcm9$ 
field of view and orientation of the {\it Chandra} ACIS-I image with the 
other data sets that we will use in this 
and subsequent papers:
(i) the $9\farcm1\times 9\farcm1$ VLT ISAAC mosaic of Pillars 1--4,
(ii) the HST WFPC2 image of Pillars 1--3 (shown in Fig.~1), and
(iii) the {\it HST} ACS mosaic of Pillar 5. 
The \citet{Sugitani2002} and \citet{TSH2002} near-infrared survey
regions are approximately coincident with the WFPC2 field.
Fig.~3 also illustrates the \citet{Belikov1999} $3\sigma$ core and corona
of NGC 6611 (dashed circles) overlayed on an optical image showing the
emission nebula, the optically bright OBA stars in the cluster core 
and the molecular clouds surrounding the cluster core.

\subsection{Astrometric Correction}

A full-resolution 2800$\times$2800 pixel image ($0.492\arcsec$ pixels),
containing ACIS CCDs I0-I3 was generated to search for X-ray sources. 
We selected 98 2MASS sources with a single obvious counterpart on the
{\it Chandra} images to register the {\it Chandra} images
on the 2MASS reference frame. The 2MASS J2000 positions and {\it Chandra}
physical pixel positions were used to derive a 4-coefficient plate scale
solution using the Starlink program ASTROM. The resulting RMS offset was less
than $0.4\arcsec$.
We also used ASTROM to correct the VLT astrometry by deriving
4- and 6-coefficient plate scale solutions from a set of 2MASS J2000
positions and VLT image pixel positions. The 4- and 6-coefficient
solutions produced RMS offsets of $0.16\arcsec$ and $0.18\arcsec$,
respectively. 

\subsection{X-ray Source Detection}

The CIAO wavelet-based tool WAVDETECT was used to detect X-ray sources
on the full-resolution 0.5-7 keV ACIS-I image. WAVDETECT scaling factors of
2, 4, and 8 were used with a false alarm probability of $10^{-5}$. 
Our criterion for source detection was 6 raw counts. A number
of spurious detections were found along the image boundaries because the 
background inside the field of view far exceeds the background in unexposed
regions. These detections were deleted after visual inspection of the
{\it Chandra}, 2MASS and VLT images. In addition, 29 regions not found
with WAVDETECT contained more than six net counts but were clearly associated
with a relatively bright 2MASS counterpart. In a few cases, WAVDETECT 
did not resolve sources that were clearly resolved in the ACIS data.
These sources were separated and added to the WAVDETECT list, resulting
in 1101 X-ray detections.

The positions output from WAVDETECT were used
to extract event files, ancillary response files, and response matrices
for each detected source. The net counts are the raw counts minus the
estimated background. We used the area of the source region and a background
count rate of 0.17 counts~s$^{-1}$ per ACIS-I CCD, as determined from a series
of local background subtractions and from source-free, blank-sky VFAINT data
from the {\it Chandra} X-ray Center. The typical background is
$1-4$ counts per source near the center of the FOV where the PSF is small,
and as high as 10 counts near the edge of the FOV. 
A local background subtraction was not used because of the high number of
resolved and unresolved sources near the cluster center.

Each source's ancillary response file is a list of effective area
versus energy at the source's position on the detector. The absorbed X-ray
flux at Earth is,
\begin{equation}
f_{\rm X} = \frac{1}{t} \sum_{i=1}^{n} \frac{e_i}{a_i(e)} - f_{\rm bkg}A,
\end{equation}
where $e_i$ is the energy of the ith photon, $a_i(e)$ is the corresponding
effective area, $t=77126.02$~s is the livetime of the observation,
A is the area of the source extraction region, and
$f_{\rm bkg}=3.605\times10^{-18}$~ergs~cm$^{-2}$~s$^{-1}$~pixel$^{-1}$
is the 0.5-7.0~keV background flux as determined from a deep, source-subtracted
observation.

The event files were used to calculate 
the mean photon energy $\bar{e} = \frac{1}{n}\Sigma_i e_i$,
and the Kolmogorov-Smirnov statistic, KS,
\begin{equation}
{\rm KS} = \sqrt{n}\,\sup|f_{i}(t)-f_{0}(t)|,
\end{equation}
where $n$ is the number of events, $e_i$ is the photon energy as measured 
by the number of accumulated electrons in the ACIS event island,
$f_i(t)$ is the normalized observed
cumulative distribution and $f_0(t)$ is the normalized model cumulative
distribution (in this case, a constant flux)\citep{Babu1996}.
When ${\rm KS} \gtrsim 1.0$, the count rate is not constant, usually
indicating flare-like activity in late-type stars.

\subsection{Source list}

The 78-ks {\it Chandra} observation of M~16 and NGC 6611 has revealed
1101 X-ray sources in the $16\farcm9\times16\farcm9$ ACIS-I field of view.
Table~1\footnote{The complete machine-readable version of Table 1 is
available in electronic format. The printed version of Table~1 lists
only those X-ray sources located within the HST WFPC2 field of view.} 
lists for each source the X-ray source number, the J2000 IAU designation,
R.A. and Declination (J2000) in degrees, raw ACIS-I counts,
background-subtracted net counts, $1\sigma$ net error, absorbed X-ray flux, 
Kolmogorov-Smirnov time variability statistic (Eq.~2), and mean photon energy.
Most of the X-ray sources are concentrated within $3\farcm6$ of the optical
center of NGC 6611 centered at
$\alpha = 18^{\mathrm h}18^{\mathrm m}40^{\mathrm s}$, 
$\delta = -13^\circ47\farcm1$ (see Fig.~3 and \citet{Belikov1999}).

Figure~4 shows a $4\arcmin\times4\arcmin$ portion of the
{\it Chandra} ACIS-I field centered on the O9.5 star BD-13~4930 and including
Pillars 1, 2 and 3.
The RGB image is color coded to indicate X-ray hardness:
0.5-1.5 keV (red), 1.5-2.5 keV (green) and 2.5-7.0 keV (blue).
The bright reddish sources like BD-13~4930 in Fig.~4 are relatively
soft OB stars seen through modest extinction,
while the white sources are nearly all low-mass T Tauri stars, which are
moderately hard X-ray sources seen through moderate extinction. The blue
sources like M16ES-1 are hard X-ray sources seen during a strong, 
very hot flare,
and/or seen through substantial column density.
Figure~5 is an optical CCD image obtained at the KPNO 0.9-m telescope
showing Pillar 4 and HH216 in [\ion{S}{2}] (red), H$\alpha$ (green) and
[\ion{O}{3}] (blue). The $3\sigma$ X-ray error circles are overlayed in red 
(black in the black and white figure).
Faint and off-axis {\it Chandra} sources generally have larger error circles.
Fig.~6 is the VLT ISAAC $K_{\rm s}HJ_{\rm s}$ RGB image of the same
field showing very good correspondence between {\it Chandra} and VLT sources.
Essentially all of the {\it Chandra} sources have near-infrared
counterparts located close to the center of the {\it Chandra}
error circle.
 
\subsection{Source Matching}

The 1101 {\it Chandra} sources in Table~1 were matched to the database of
infrared sources described in \S2. Specifically, the {\it Chandra} field
of view contains 7970 sources in the 2MASS Point Source Catalog with reliably
measured $JHK_{\rm s}$ photometry. A {\it Chandra} source was matched 
to the nearest
2MASS source within $2\arcsec$, although {\it Chandra}'s resolution of 
less than
$1\arcsec$ is smaller than the nominal 2MASS resolution of $4\arcsec$.
Source \#5 is located near the western edge of the ACIS chip, $5\farcs0$
from source 2MASS 18181067-1349453. This match was added by hand because of
the uncertainty of the {\it Chandra} position. This source does not
appear to be associated with the nearby optical star Belikov 31864.
The RMS offsets between {\it Chandra} and 2MASS for 812 matches are
$\Delta\alpha=0.36\arcsec$ and $\Delta\delta=0.38\arcsec$.

The {\it Chandra} error circles ($3\sigma$ error radius) were
overlayed on the VLT image. The XY coordinates of any matching VLT sources
were used to extract aperture photometry in all 3 ISAAC bands.
The RMS offsets between {\it Chandra} and VLT for 550 matches are
$\Delta\alpha=0.34\arcsec$ and $\Delta\delta=0.40\arcsec$.

Table~2\footnote{The complete machine-readable version of Table 2 is available
in electronic format. The printed version of Table~2 lists only X-ray 
sources and infrared sources located within the HST WFPC2 field of view.}
lists the X-ray source number, IAU designation, VLT
R.A., Decl., and offset from the {\it Chandra} position, 
VLT $J_{\rm s}HK_{\rm s}$ photometry, 2MASS
R.A., Decl., and offset from the {\it Chandra} position, 
and 2MASS $JHK_{\rm s}$ photometry.
Stars in Table~2 with 2MASS data but no VLT data are
outside the $9\farcm1\times9\farcm1$ VLT FOV
or are saturated in the VLT images.
Also listed are the catalog number, $B$ and $V$ magnitudes
of \citet{Belikov1999}, and the spectral type and visual
absorption $A_V$ from \citet{Hillenbrand1993}.
In three cases, a single {\it Chandra} source has four VLT counterparts,
in nine cases a single {\it Chandra} source has three VLT counterparts,
and in 31 cases a single {\it Chandra} source has two VLT counterparts.
The other stars matching an X-ray source are listed directly below the
primary match and have no data for the {\it Chandra} source number in column 1.

Of the 1101 X-ray sources in the electronic version of Table~2
(see Paper~II),
550 are matched to one or more VLT $K_{\rm s}$-band sources,
812 are matched to one 2MASS $K_{\rm s}$-band source, and only
85 are matched to the \citet{Belikov1999} or \citet{Hillenbrand1993}
catalogs. Overall, 968 of 1101 are matched with at least one VLT or 
2MASS source.

\section{Results}

In this paper we focus our attention on the young stellar sources and EGGs in 
and near Pillars 1 through 4. In subsequent papers of the series we will 
address other aspects of this rich data set.

\subsection{YSOs Near Pillars 1, 2 and 3}

{\it Chandra} detected forty X-ray sources (see Table 3) in the 
$2\farcm5\times3\farcm0$ field of \citet{Sugitani2002}
centered on Pillars 1, 2 and 3. The Sugitani field overlaps and is
slightly larger than the WFPC2 field (see Fig.~4).
In the next section we describe our method
for determining the intrinsic luminosities and other properties 
of these sources.
To better identify the nature of these sources and whether they are likely
cluster members or instead foreground or background sources, we plot in
Figure~7 the $JHK_{\rm s}$ color-color diagram for the 784 stars in the
\citet{Sugitani2002} field including Pillars 1, 2 and 3. The 40
{\it Chandra}-detected stars in Table~3
are shown as large filled symbols. Large circles represent bright stars with
$J \leq 17$; dots represent fainter stars. Also shown is the locus
of dereddened ZAMS stars (dash-dot line), the locus of dereddened
classical T Tauri stars (dash-triple dot line), reddening lines
for cool giants and dwarfs (dashed lines), and an $A_V=10$~mag
extinction vector (solid arrow).
We identify the following stellar populations in the color-color diagram 
(Figure~7).

\begin{description}

\item[(1)] At\footnote{We use the symbol $\Delta$ (color excess) to refer to 
redenning produced by a disk, and the symbol $E$ (color excess) to refer to 
redenning produced by the different interstellar attenuation in the 
two passbands.} $\Delta (H-K_{\rm s}) \gtrsim 1$ there 
is a small group of reddened
stars with large infrared excesses. Of the six ``P'' sources in this portion
of the color-color diagram proposed by \citet{Sugitani2002} as Class I
protostars, {\it Chandra} only detected P1=M16ES-1 with $\log L_{\rm X}=32.2$.
The other sources were not detected with an intrinsic X-ray luminosity
threshold $\log L_{\rm X} < 29.8$.

\item[(2)] To the right of the dashed redenning lines, close to or above the
dereddened CTTS line, is a small population of CTTSs (labelled CT) 
with large infrared 
excesses (or Herbig Ae/Be stars with small excesses). These are the ``T'' 
sources of \citet{Sugitani2002}. If we only consider stars with $J<17$ and
$J-H<1.5$, then {\it Chandra} detected 7 out of 21, including the 
intermediate-mass T Tauri star T1=M16ES-2.

\item[(3)] At $J-H \lesssim 1.1$ and between the dashed redenning lines
we see a larger population of lightly reddened ($1<A_V<7$) stars with
nearly photospheric colors. Most of these are naked T Tauri stars (labelled NT)
and some classical T Tauri stars with small $H-K_{\rm s}$ excesses.
{\it Chandra} detected 25 out of 60 of these sources with $J<17$ above the 
threshold $L_{\rm X}\approx 29.8$.

\item[(4)] {\it Chandra} detected the O9.5 star 29726=BD$-13\arcdeg 4930$,
the hot star located at the base of Pillar~1 at 
$(H-K_{\rm s},J-H)=(0.04,-0.02)$. This star shows very little redenning with 
$\log L_{\rm X}=30.8$, a factor of 30 lower than very X-ray luminous stars
M16ES-1 and $\theta^1$~Ori~C (see Table~4). BD$-13\arcdeg 4930$ is the only 
O star near Pillars 1, 2, and 3. It and the high-mass YSO M16ES-1 are 
the most luminous X-ray sources of the 40 listed in Table~3. 
BD$-13\arcdeg 4930$ has a low K-S value indicating no detectable
time variability and the lowest mean photon energy of the sources listed in
Table~3. This suggests that the X-ray emission from this star is like that
of most other O stars and early B stars, with $kT\sim 0.5$~keV and
$\log L_{\rm X}/L_{\rm bol} \sim -7$ \citep{Owocki1999}.

\item[(5)] At $J-H \lesssim 0.7$ we see a small population
of relatively bright, unreddened ($A_V<1$), presumably foreground 
stars (labelled FS),
only three of which are detected with {\it Chandra}. Our reason for
identifying these stars as foreground sources is based on the following
argument. The population
of ``blue'' bright stars has a low X-ray detection fraction.
Since T~Tauri stars are very X-ray active well into their 
main-sequence lifetime, low-mass members of NGC~6611 cannot be both
optically bright and X-ray faint. However, these stars could be
intermediate-mass stars close to the main sequence (spectral type B3--F5).
At 2--3~Myr, few of these stars would have infrared excesses from disks.
They are expected to be X-ray faint because they do not have large 
outer convective zones to drive a magnetic dynamo and weak winds
incapable of accelerating X-ray emitting shocks.

\item[(6)] A large number of highly reddened ($A_V>7$) stars
lie close to the upper reddening line; none of these stars are
detected with {\it Chandra}. We believe that these stars are 
background giants (labelled BG) in the Galaxy redenned 
by the cluster gas and dust.
They cannot be foreground stars because the number of stars in this part 
of the infrared color-color diagram is far too great. They are unlikely to
be M~16 cluster members because unlike true cluster members 
they do not emit X-rays,
do not have disks, and are highly obscured. It is more likely that they
are located behind M~16 since they have much larger $A_{\rm V}$ 
than the X-ray bright stars, even allowing for the X-ray bias toward low
$A_{\rm V}$ stars. Since M~16 lies in the Galactic plane, stars from 2--8~kpc
are in the line of sight. Although giants are less numerous in the
disk than dwarfs, they are 100--1000 times more luminouis than dwarfs 
of the same effective temperature. Therefore, one detects more giants than
dwarfs. In the Galactic plane the giant clump occurs at J-H=1.0, but the
stars behind M~16 have J-H $\approx 2.0$. This suggests that the average 
visual absorption through M~16 is $<A_{\rm V}> \approx 10$. We note that
in the line of sight through the pillars $A_{\rm V}$ exceeds 25.

\end{description}

\subsection{X-ray Luminosities}

To determine accurate X-ray luminosities, X-ray spectra are
usually fit to an absorbed thermal model, the critical parameters affecting
$L_{\rm X}$ being hydrogen column density, plasma temperature,
emission measure, and metal abundance. When the number of counts in an X-ray
spectrum drops below a few hundred, the fit parameters and $L_{\rm X}$ become
highly uncertain. We employ a method discussed by \citet{Gagne2004} for 
determining $L_{\rm X}$ for X-ray faint YSOs with relatively accurate
near-infrared photometry.  Here we consider only the 40
X-ray sources in Table~3 in the \citet{Sugitani2002} field around 
Pillars 1, 2 and 3.

To estimate $L_{\rm X}$, we proceed by two steps:
(1) We first estimate the infrared excess $\Delta(J-H)$,
$A_V$, and $N_{\mathrm H}$ for each source and (2) we then estimate 
the absorbed flux and $L_{\mathrm X}$ assuming $d\approx2$~kpc.
In the first step, stars between the dashed redenning vectors in Fig.~7
(the ``NT'' sources) have low $\Delta(H-K_{\rm s})$: we assume that these 
stars have no excess $J-H$ reddening from a disk.
Typically, T Tauri stars have spectral type late-G to late-M with
intrinsic color
$(J-H)_0$ in the range 0.33 to 0.66~mag \citep{Meyer1997}.
Since such stars in M~16 do not have measured spectral types, we assume a mean
spectral type of K0 and $(J-H)_0=0.45$. For the ``T'' sources 
(labelled CT in Fig.~7) between the
dashed lines, we assume an intrinsic color $(J-H)_0=0.65$ to account for
reddening from a disk. We assume $A_V = 9.37 E(J-H)$,
where $E(J-H)=(J-H)-(J-H)_0$ \citep{RL1985} and
$N_{\mathrm H}/A_V=1.57\times10^{21}$~cm$^{-2}$~mag$^{-1}$ \citep{Vuong2003}.
We estimate that this procedure leads to an uncertainty of $0.15$~mag in
$E(J-H)$ implying an uncertainty of $\sim1.4$~mag in $A_V$ and
$\sim 2.5\times10^{21}$~cm$^{-2}$ in $N_{\mathrm H}$.
This leads to an uncertainty\footnote{The estimated errors in 
$\log L_{\rm X}$ following the procedure outlined in Appendix B
of \citet{Gagne2004} depend on the number of raw X-ray counts and the 
accuracy of the infrared photometry. For 
typical stars in our sample with $\log L_{\rm X}\approx 7.4$ and
$A_{\rm V}\approx 4$ (see Table~3), the availability of infrared photometry 
for extimating $A_{\rm V}$ and thus $N_{\rm H}$ is critically important. In 
the absence of such data, errors in $\log L_{\rm X}$ are below $\pm 0.10$ for 
sources with more than 100 raw X-ray counts but increase rapidly to 
$\Delta \log L_{\rm X} = ^{+0.43}_{-0.14}$ for 25 raw X-ray counts and to
$^{+0.88}_{-0.28}$ for 10 raw counts. By contrast, when $N_{\rm H}$ is known 
to $\pm 10$\%, the error in the inferred $\log L_{\rm X}$ is below $\pm 0.10$
even for 10 raw X-ray counts. These estimated errors in $\log L_{\rm X}$ are 
likely lower limits due to various systematic errors and errors in estimating 
$N_{\rm H}$ for those sources with faint infrared magnitudes and poorly known 
spectral types.}
of $\sim 0.13$ in $\log L_{\mathrm X}$.
For M16ES-1 and M16ES-2 we use the $A_V$ estimates of \citet{MA2002}.
The $JHK$ photometry for BD$-13\arcdeg 4930$ (O9.5~V) comes from 
\citet{Hillenbrand1993}; we assume $(J-H)_0=-0.13$ \citep{BB1988}.
For J181850.8--134842 and J181852.1--134928, the $J_{\mathrm s}HK_{\mathrm s}$
photometry was incomplete or uncertain. For these two, we assume
a median $A_V \approx 4.3$ and consider their X-ray luminosities 
to be uncertain.

In the second step, we use the mean energy $\bar{e}$
and $N_{\mathrm H}$ to estimate the flux correction factor (the ratio of
unabsorbed to absorbed X-ray flux). The absorbed flux listed in Table~1
is calculated directly from the photon event list and the effective area
as a function of energy determined at each source's position on the detector.
The absorbed flux calculated in this manner does not depend on any assumptions
about the emitter or absorber.  To determine the unabsorbed flux (i.e., the
absorption-corrected flux), we have
generated a grid of synthetic ACIS-I spectra in XSPEC 11.0 using the
ISM absorption model of \citet{MM1983} and a single-temperature
VAPEC emission model \citep{Smith2001}.  We use solar abundances 
\citep{Anders1989}
for all elements except iron, for which we assume 0.5 times solar.
Synthetic spectra are generated across a grid of column density in the
range $20.0\leq\log N_{\mathrm H}\leq24.0$~cm$^{-2}$ and temperature
$6.0\leq\log T\leq9.0$~K in $\log T$ steps of 0.1. The mean energy and
flux correction factor are calculated from each synthetic spectrum.
For each source, the flux correction factor is interpolated from the grid
using the observed values of $\bar{e}$ and $N_{\mathrm H}$.
Unabsorbed flux values obtained this way have been compared with those from
explicit fits to ACIS-I spectra of X-ray bright YSOs in $\rho$~Oph
\citep{Gagne2004} and NGC~2024 \citep{Skinner2003}.
The unabsorbed flux values usually agree to within 10\%, suggesting that
this is a robust way of estimating X-ray luminosity for faint sources.
The 0.5-7~keV X-ray luminosity is calculated from the unabsorbed flux assuming
a distance of 2~kpc \citep{Hillenbrand1993}.
For the 40 {\it Chandra} sources in the \citet{Sugitani2002} field,
Table~3 lists the source number, IAU designation {\it Chandra} source name,
near-infrared source name, offset from the {\it Chandra} position,
net counts, K-S statistic,
mean energy $\bar{e}$, absorbed X-ray flux 
(ergs cm$^{-2}$ s$^{-1}$),
$J$, $H$, $K_{\mathrm s}$ magnitudes, visual absorption $A_V$, 
and unabsorbed $\log L_{\mathrm X}$.
As shown in Table~3, our detection limit for the unabsorbed X-ray luminosity
of stars in M16 is $\log L_{\rm X}\approx 29.8$.
These quantities will be used extensively in subsequent papers of this series.

\subsection{The Massive YSO M16ES-1}

The most remarkable source in the {\it Chandra} field of view is the very
hard source M16ES-1 at the head of Pillar 1.
It is not coincident with any optically visible star or EGG (see Fig.~1).  
In the
VLT+ISAAC $J_{\mathrm s} H K_{\mathrm s}$ RGB image of Pillar 1 (Fig.~2)
\citep{MA2002},
{\it Chandra} source positions are indicated with circles.  The 113-count
{\it Chandra} source is associated with the reddened source M16ES-1 
(yellow in Fig.~2) directly
south of the reflection nebula at the head of Pillar 1.
\citet{TSH2002} estimate $L_{\mathrm bol}\approx200L_{\odot}$ by integrating
the far-, mid- and near-infrared flux within a $2\arcsec$ radius aperture.
This luminosity suggests either a zero age main sequence (ZAMS) B star, 
a massive pre-main-sequence star, or a small cluster of low-mass YSOs.
\citet{TSH2002} argue that M16ES-1 is probably not a ZAMS B star because
of the absence of Pa$\alpha$ emission. 
Based on the PMS tracks of \citet{PS1999}, M16ES-1's luminosity suggests
it is a 4--5 $M_\odot$ star. Its large infrared excess (Fig.~7) suggests a
substantial disk or an ultracompact HII region \citep{MA2002}.

Assuming $A_V\approx27$ \citep{MA2002}, we estimate that 
the 0.5--7~keV X-ray luminosity of
M16ES-1 is $L_{\mathrm X}\approx 1.6\times10^{32}$~ergs~s$^{-1}$. 
M16ES-1 was not flaring (KS = 0.48) during the {\it Chandra} observation, but  
its mean photon energy was very large, $\bar{e}=3.3\pm0.2$~keV.
Assuming $A_V=27\pm2$~mag, we find its plasma temperature to be 
$2.2^{+1.0}_{-0.6}$~keV.
$L_{\mathrm bol}\approx200L_{\odot}$ \citep{TSH2002} implies that
$L_{\mathrm X}/L_{\mathrm bol}\approx2.1\times10^{-4}$.
This high plasma temperature and large $L_{\mathrm X}/L_{\mathrm bol}$,
generally require magnetic confinement and
are often seen on active, magnetically heated
coronal sources like Class I protostars and Class II and III T Tauri stars.
However, the X-ray luminosity of M16ES-1 exceeds that of any single nonflaring,
low-mass YSO in Orion \citep{Gagne1995,Getman2005a} or
$\rho$~Oph \citep{Imanishi2001,Gagne2004}.
To our knowledge,  O-type stars are the only young stars with steady X-ray
emission exceeding $\log L_{\rm X}\approx 32$. The only other high-mass
star known to have such a hard X-ray spectrum is the magnetic O7~V star
$\theta^1$~Orionis~C \citep{Gagne2005}, the ionizing source of the 
Orion Nebula (cf. Feigelson et al. 2005). \citet{Gagne2005} explain the hard 
X-ray spectrum and narrow line widths of this 
star with a magnetically channeled wind shock (MCWS) model in which the X-rays 
are produced in strong shocks when the magnetically channeled winds from each 
hemisphere meet near the magnetic equator.

In Table~4 we compare the properties of M16ES-1 to two other young stars
($\theta^1$~Ori~C and Oph~S1) that are known to be magnetic and two 
somewhat older stars ($\zeta$~Ori and $\tau$~CMa) that show little or 
no evidence for
magnetic fields. $\theta^1$~Ori~C has a measured magnetic field 
\citep{Donati2002}, and the presence of a 
magnetic field in Oph~S1 is inferred from its
nonthermal gyrosynchrotron emission \citep{Andre1988}. \citet{Waldron2000} 
suggest that $\zeta$~Ori may have a weak magnetic field because the
helium-like Si~XIII line ratios are consistent with the bulk of 
the Si~XIII emission being produced in
magnetic loops located near the base of the wind rather than in shocks 
embedded in the wind.  There are no direct measurements of magnetic fields
or nonthermal gyrosynchrotron radio emission from $\zeta$~Ori or $\tau$~CMa. 

\citet{Schulz2003} and others have argued that a clear indicator of the
presence of coronal magnetic fields is the presence of hard X-ray emission
indicative of substantial amounts of coronal plasma 
hotter than 1 keV (12.4 MK).
In their analysis of {\em Chandra} ACIS-S/HETG spectra of the Orion Trapezium 
stars, including $\theta^1$~Ori~C,
they find that most of the emission is at temperatures above 20~MK 
with a small component at temperatures below 10~MK. 
For these young stars (typical ages
0.3 Myr), the emission measure distributions show a prominent gap at 7.15 MK
(0.9 keV). They ascribe the weak low temperature emission to shocks in the 
stellar winds and the high temperature emission to hot magnetically heated 
plasma confined in coronal magnetic loops. The narrow X-ray emission lines
with no Doppler shifts, which are seen in $\theta^1$~Ori~C and the 
other Trapezium stars, are consistent with most of the emission produced
in a magnetically confined plasma. On the 
other hand, broad blue-shifted X-ray emission lines are commonly observed in
the older O stars and are usually explained by the
line-force instability wind shock model with the shocks located well out
in the expanding wind \citep{Owocki1988}. 
For these stars the soft X-ray emission from
the wind is characterized by 
$L_{\rm X}^{\rm wind}/L_{\rm bol}\approx 10^{-7.2}$.
We list in Table~4 the total X-ray luminosity, $L_{\rm X}^{\rm total}$ and the
X-ray luminosity from the wind, $L_{\rm X}^{\rm wind}$ either from
\citet{Schulz2003} or computed from 
$L_{\rm X}^{\rm wind}/L_{\rm bol}\approx 10^{-7.2}$.
The data in Table~4 show that M16ES-1 has high 
coronal plasma temperature and total X-ray luminosity far larger than 
$L_{\rm X}^{\rm wind}$, which is similar to the two 
young stars with known magnetic fields ($\theta^1$~Ori~C and Oph S1) 
but very different from the two older stars with no evidence 
for magnetic fields ($\tau$~CMa) or perhaps weak magnetic fields ($\zeta$~Ori).
We conclude therefore that M16ES-1 is most likely
a magnetically active, high-mass YSO for which the hot plasma may be 
heated as in the MCWS model. Since
the X-ray emitting regions are likely magnetically heated, longer 
duration X-ray observations may reveal evidence of magnetic breakout events.

\subsection{The T Tauri Star M16ES-2}

The weakest {\it Chandra} source we list is the intermediate-mass
T Tauri star M16ES-2 located near the tip of Pillar 2 (see Figs.~1 and 2).  
With only 5 raw counts (4.7 net counts after background
subtraction), it falls below the 6-count cutoff applied to the general
source population.  We include it in Table~3 because those 5 counts are 
spatially coincident with M16ES-2. $\log L_{\mathrm X}\approx 30.1$ 
for this source is close to the median for 2-$6M_{\odot}$
stars in Orion \citep{Feigelson2002}. Assuming for the bolometric luminosity 
$L_{\mathrm bol}\sim 20 L_{\odot}$ \citep{TSH2002}, we find that 
$\log L_{\mathrm X}/L_{\mathrm bol} \sim -4.8$, which is rather low 
for T Tauri stars. This ratio could be low because of the errors associated 
with the small number of detected counts or perhaps the star is nearly as young
as the EGGs, which are not detected X-ray sources (see \S3.6).

\subsection{HH216}

Another weak source is associated with the Herbig-Haro object HH216,
seen as a bright H$\alpha$ bow shock in Fig.~5 and as a wisp of bluish
$1.2\micron$ emission in Fig.~6.
The {\it Chandra} feature consists of 8 counts centered at
$\alpha=18^{\mathrm h}18^{\mathrm m}55\fs37$,
$\delta=-13\arcdeg51\arcmin45\farcs6$ (J2000)
in a line parallel to the bow shock of HH216.
The X-ray counts appear to be $\sim 1\arcsec$
directly behind (southeast) of the K-band emission peak.
The X-ray feature is approximately $1\arcsec$ (two pixels) wide in the EW
direction and $4\arcsec$ (8 pixels) long in the NS direction.
The mean photon energy is $1.9\pm0.6$~keV. Assuming these counts were 
produced by the HH216 shock at a distance $d\approx2$~kpc, this source has 
$\log L_{\mathrm X}\approx 30.0$, comparable to
HH2 \citep{Pravdo2001}, but less luminous than HH 80 and 81
\citep{Pravdo2004}, with $\log L_{\mathrm X}\approx 31.6$
and an estimated plasma temperature $kT\approx 0.13$.

In Pillar~4, directly south of Pillars 1--3,
\citet{Andersen2004} identify two highly absorbed sources of
extended emission that appear to be aligned with HH216. These
appear as red knots of $2.2\micron$ emission $\sim 6\arcsec$
southeast and northwest of the unseen exciting source of both
jets and HH216 in Fig.~6. Alternatively, the exciting source could be
located closer to the red nebulosity at the tip of the pillar.
Neither the jets nor the exciting source
are visible optically in Fig.~5. Although 4 ACIS
counts are aligned along the northernwestern jet (the jet closest to HH~216)
at $\alpha=18^{\mathrm h}18^{\mathrm m}59\fs0$,
$\delta=-13\arcdeg52\arcmin48\arcsec$ (J2000),
we do not consider this to be a statistically significant detection, due to the
small number of counts detected off-axis.
The likely driving source for HH216, identified by \citet{Andersen2004} in
their C$^{18}$O image, is Core C centered at
$\alpha = 18^{\mathrm h}18^{\mathrm m}59\fs25$,
$\delta = -13\arcdeg53\arcmin00\arcsec$ (J2000) with a radius of $25\arcsec$.
We find no X-ray source at this location, perhaps because the source is
heavily obscured.  The southeastern jet at
$\alpha=18^{\mathrm h}18^{\mathrm m}59\fs36$,
$\delta=-13\arcdeg52\arcmin56\farcs2$ (J2000)
is not detected with {\it Chandra}. We note that \citet{Favata2002}
detected X-rays from the jet emanating from the L1551~IRS5 protostar
with a temperature of about 4MK and $\log L_{\rm X} \approx 29.5$, about
a factor of 2 below our detection threshold. 
\citet{Favata2006} also detected X-rays from the HH~154 jet source.
The X-rays are likely produced at the working surface where the jet impacts
the circumstellar medium.

\subsection{Evaporating Gaseous Globules in Pillars 1, 2 and 3}

To search for X-ray emission from the evaporating gaseous globules (EGGs) 
discovered by \citet{Hester1996}, the {\it Chandra} image was 
registered to the HST WFPC2
H$\alpha$ image using the bright stars BD$-13\arcdeg 4930$ and CDS~975. 
The locations of the 73 EGGs identified by \citet{Hester1996} are indicated in
Fig.~1 as diamonds, but none of the EGGs 
are located inside the {\it Chandra} $3\sigma$ position uncertainty 
circles\footnote{The X-ray position uncertainty circle
for source M16ES-2 at the tip of Pillar 2 includes near its edge three
diamonds. This error circle is very large because the X-ray source is very weak
(5 raw counts). Since the IR point source identified as M16ES-2 is more than 
$2\arcsec$ from the three EGGs, we do not believe that these EGGs 
are X-ray sources.}. To verify this result,
the {\it Chandra} 0.5--7~keV image was registered with the VLT ISAAC K-band
image using 48 common stars (Fig.~2): the final fit yielded an RMS error of 
$0\farcs3$.
The HST/VLT image registration of \citet{MA2002} had an RMS
error of $0\farcs02$. Although 11 of the 73 EGGs have an associated
infrared point source likely indicating the presence of an embedded YSO, 
none of these infrared sources has an X-ray counterpart
located within $2\arcsec$.

Seven of the 11 infrared EGG sources appear to be substellar \citep{MA2002}.
The nondetection of these sources by {\it Chandra} is not suprising,  
because of the 8 spectroscopically identified brown dwarfs detected as 
X-ray sources in the COUP 9.6 day {\em Chandra} ACIS-I exposure of the ONC,
the ``quiescent'' X-ray luminosities are in the range 
$\log L_{\rm X} =$ 27.3--28.5 \citep{Preibisch2005a}. The ONC brown 
dwarfs flared at the rate of one large flare 
per 650~ks per object. If these stars are
representative of PMS brown dwarfs embedded in the EGGs, then during the 
78~ks observation of M~16 one flare increasing $L_{\rm X}$ by roughly 
a factor of 10 \citep{Preibisch2005a} could have occured. This flare would 
have been below our detection threshold of $\log L_{\rm X} \approx 29.8$.
Therefore, if the substellar infrared sources in the EGGs emit X-rays, 
then their X-ray luminosities could be larger than the ONC brown dwarfs
and not be detected in our 78~ks observation of M~16.

The other four near-infrared EGG sources (E25, E31, E35, and E42) were also not
detected with {\it Chandra}. They range in mass from 0.35 to $1.0 M_{\odot}$.
In the COUP ONC survey of YSOs with masses 0.5--0.9$M_{\odot}$,
90\% of the stars have $\log L_{\rm X}$ in the range 29.3--31.2 with the 50\%
point in the cumulative X-ray luminosity function at 30.4. The corresponding
X-ray luminosities for the 0.1--0.5$M_{\odot}$ range are 28.3--30.4 with the
50\% point at 29.6 \citep{Preibisch2005b}. Since our detection limit was
$\log L_{\rm X}\approx 29.8$, we should have detected at least two of the 
three EGGs with masses greater than 0.5$M_{\odot}$ if the EGGs contain stars
similar to ONC stars of similar mass. Could the intermediate mass EGG 
sources have escaped
detection because of large extinction? \citet{MA2002} list visual extinctions
of $A_{\rm V}$ = 4--22 for these four stars, but the most 
massive EGG point source 
(E42 with $1.0M_{\odot}$) has only $A_{\rm V} = 4$ mag. The X-ray extinction
for this presumably hard source should not be large enough to prevent its
detection if the point source in E42 were as X-ray luminous as a typical ONC 
PMS star of similar mass. We conclude therefore that the 0.35--1.0$M_{\odot}$ 
point sources in the 
EGGs either do not emit X-rays or their X-ray luminosities are far 
smaller than the PMS stars in the ONC with similar masses.

To determine the likelihood that the EGGs are less X-ray luminous than the
ONC YSOs of similar mass in a more quantitative manner, we simulated 
ACIS-I spectra in XSPEC assuming 
$N_{\rm H}/A_{\rm V} = 1.57\times 10^{+21}$ cm$^{-2}$/mag \citep{Vuong2003}.
The APEC model parameters were $N_{\rm H}$ (units $10^{+22}$ cm$^{-2}$)
from $A_{\rm V}$,
$kT = 2.0$ keV, and $Z=0.5$ solar which is typical for YSOs. The 
6-count upper limit, 77.11~ks exposure, and assumed model parameters then
allowed us to compute upper limits for the observed flux $f_{\rm X}$ in the
0.5--7 keV band and the flux corrected for absorption $f_{\rm X}^{\rm corr}$,
both in units $10^{-16}$ ergs cm$^{-2}$ s$^{-1}$. These quantities and the 
corresponding source X-ray luminosities $L_{\rm X}$ (units 
$10^{30}$ ergs s$^{-1}$) are listed in Table~5 for the four intermediate 
mass EGGs studied by \citet{MA2002}.

We also list in Table~5 the approximate fraction (frac) of YSOs close in mass
to the four EGGs
with $L_{\rm X}$ above this limit in the COUP survey. The number of 
expected detections is the sum of the fractions: 2.28. Our result that 
{\it Chandra} detected none of the four EGGs is statistically
significant. The likelihood of detecting none of these four EGGs is only 1.5\%,
a $2.4\sigma$ result. Although N=4 is small, we conclude that the
EGG point sources do not emit X-rays at the level that one would expect
from T Tauri stars. Add to this that none of the other EGGs were detected,
and we are left with the conclusion that either 
(a) the EGGs do not form stars, or
(b) the YSOs in the EGGs have not yet become X-ray active. 

Another EGG, E23, is notable because a faint jet is seen emanating from it
in the J-band image of \citet{MA2002}.  E23 has no near-infrared
counterpart and no {\it Chandra} counts are detected near E23 or its jet.
While deeper X-ray observations are needed,
it is clear that unlike T Tauri stars and other YSOs, EGGS and their 
embedded cores as a class are not strong X-ray emitters. This could result from
either their intrinsic X-ray emission being weak (perhaps because the 
cores have not yet developed organized magnetic fields)
or substantial extinction in the line of sight to the
X-ray emission region. The nondetection of E42 and the three other EGGs with 
intermediate mass embedded YSOs, suggests that these sources, which are likely 
examples of the very youngest stage of star formation, do not have the 
ability to be strong X-ray emitters.

With the notable exception of M16ES-1 and M16ES-2, the X-ray source
population is not clustered in or near the columns of gas and dust.
This is in stark contrast to active star-forming regions like NGC 2024
where the {\it Chandra} data reveal a dense, optically invisible cluster
of hard, flaring X-ray sources \citep{Skinner2003}.
The pre-protostellar cores at the tips of the columns described 
by \citet{Pound1998} and \citet{White1999} are not (yet)
X-ray emitting YSOs.  These cold
cores, the infrared EGG sources, and their slightly older siblings
M16ES-1 and M16ES-2 appear to be the last vestiges of star formation
in the M~16 molecular cloud.

\section{Summary and Conclusions}

As a result of its high angular resolution and sensitivity,
{\it Chandra} is the instrument of choice for X-ray imaging of crowded fields.
Our 78~ks {\it Chandra} ACIS-I image of the M~16, 
the first X-ray image obtained of the Eagle Nebula region,
has revealed 1101 X-ray sources, nearly all of which are members of M~16 or 
NGC 6611. This $17^{\prime}\times 17^{\prime}$ ACIS-I image contains one
of the densest concentrations of X-ray sources ever observed.
For the X-ray and infrared sources near Pillars 1--4 in M~16 we find:

\begin{enumerate}

\item The near-infrared color-color diagram of the M~16 pillars
shows that the vast majority of X-ray sources 
are moderately reddened Class III YSOs or Class II YSOs with small
infrared excesses. Most of these T Tauri stars belong to the
1--6 Myr-old NGC 6611 star cluster.

\item A relatively small number of hard X-ray sources in the ACIS field 
are deeply embedded protostars located in or near the
``Pillars of Creation'', the well-known dark molecular clouds in M~16.
These younger stars represent a more recent stage of star formation,
possibly induced by photoevaporation and photoionization 
of the dark cloud by the nearby O stars.

\item The most luminous X-ray source ($\log L_{\mathrm X} = 32.2$) is M16ES-1,
also known as YSO~1 and P1,
located at the tip of Pillar 1 facing the illuminating O stars.
It was not flaring during the {\it Chandra} observation.
The large value of $L_{\rm X}/L_{\rm bol} \approx 2.1\times 10^{-4}$ and
very high mean photon energy, $\bar{e} = 3.3\pm 0.2$~keV, of M16ES-1 are 
X-ray properties that are very different from typical O stars for which the
X-rays are produced in weak shocks in their line-force unstable winds.
Since the X-ray properties of M16ES-1 are similiar to those of
the well studied magnetic O7~V star $\theta^1$~Orionis C (see section 3.3), 
the ionizing source of the Orion Nebula, we propose that
M16ES-1 is another example of an O star with 
heating by magnetically channeled wind shocks.

\item A most interesting aspect of M~16 is the presence of  
``evaporating gaseous globules'' (EGGs), deeply embedded infrared sources 
located at the edges of the pillars. \citet{Hester1996} and \citet{Hester2004}
have argued that the EGGs are a very early stage in protostellar evolution when
shock waves and photoevaporation of the molecular gas in the pillars by the 
harsh radiation from the nearby O stars produce a dense core 
that could become a star. We find that none of the 73 EGGs studied in the 
near-infrared by \citet{MA2002} are X-ray sources above our 
detection threshold. 
Eleven of the EGGs have faint near-infrared point sources. 
Of these, seven have substellar masses and are not expected to have X-ray 
emission above our detection threshold. The nondetection (above our threshold) 
of the four EGGs with core masses in the 
range $0.35-1M_{\odot}$ indicates that either (a) the EGGs do not contain YSOs 
or (b) that at a very early stage of evolution YSOs have not yet become 
X-ray active.

\item There is a weak X-ray source associated with the Herbig-Haro object 
HH216 with a mean photon energy of $1.9\pm 0.6$~keV. Assuming that the X-rays 
are produced by the HH216 shock, the source has $\log L_{\rm X}\approx 30.0$,
comparable to the X-ray luminosity of HH2 in Orion.  
The exciting source for HH216 was not detected in X-rays.

\end{enumerate}

Future papers in this series will focus on matching the X-ray 
and near-infrared sources
in M~16 with mid-infrared sources detected with the IRAC camera on the
{\it Spitzer} Space Telescope as part of the GLIMPSE survey to
determine the mass, age and disk properties of the YSOs.

\acknowledgements

This work is supported by NASA through grant H-04630D to the University of 
Colorado. This project has made extensive use of the Simbad and 2MASS 
databases administered by the Centre de Donn\'ees Astronomiques in Strasbourg,
France and the Infrared Processing and Analysis Center in Pasadena, CA.
This paper presents data obtained with the {\it Chandra} X-ray Observatory,
the {\it Hubble} Space Telescope, the Kitt Peak National Observatory 0.9-m
Telescope and the European Southern Observatory Antu Very Large Telescope.
The HST image was created by Jeff Hester (Arizona State University), courtesy 
of NASA and the Space Telescope Science Institute. The KPNO image was 
created by Travis Rector (University of Alaska), courtesy of NOAO and AURA. 
We thank the 
anonymous referee for his/her very thoughtful and useful suggestions, and we
thank Kosta Getman for providing data on the fraction of YSOs in the COUP
survey with X-ray luminosities above our threshold.

\clearpage
\begin{deluxetable}{lcccrrrccc}
\tabletypesize{\scriptsize}
\tablewidth{0pt}
\tablecolumns{14}
\small
\tablecaption{Chandra X-ray sources in M~16: WFPC2 Field}
\tablenum{1}
\tablehead{
\colhead{Source} &
\colhead{Designation} &
\colhead{R.A.} &
\colhead{Decl.} &
\colhead{Raw} &
\colhead{Net} &
\colhead{$\sigma$} &
\colhead{$10^{16}$ $f_{\rm X}$} &
\colhead{KS} &
\colhead{$\bar{e}$} \\
\colhead{} &
\colhead{} &
\multicolumn{2}{c}{(J2000)} &
\multicolumn{3}{c}{(counts)} &
\colhead{(ergs cm$^{-2}$ s$^{-1}$)} &
\colhead{} &
\colhead{(keV)} }
\startdata
 700 & J181847.4-135043 & 18 18 47.48 & -13 50 43.3 &    14 &    13.5 &     4.8 & 16.9 &   0.94 &   2.04 \\
 705 & J181847.6-134903 & 18 18 47.62 & -13 49 03.6 &   115 &   114.6 &    11.8 & 97.9 &   0.85 &   1.69 \\
 710 & J181847.8-135056 & 18 18 47.84 & -13 50 56.8 &    11 &    10.7 &     4.4 & \phn6.0 &   0.49 &   1.41 \\
 717 & J181847.9-135058 & 18 18 47.97 & -13 50 58.3 &     6 &     5.8 &     3.6 & 10.5 &   0.97 &   2.71 \\
 725 & J181848.2-134908 & 18 18 48.22 & -13 49 08.1 &   120 &   119.2 &    12.0 & 109.9\phn &   0.87 &   1.68 \\
 726 & J181848.2-134858 & 18 18 48.23 & -13 48 58.2 &    25 &    24.6 &     6.1 & 38.6 &   0.61 &   2.04 \\
 728 & J181848.3-134914 & 18 18 48.35 & -13 49 14.6 &    28 &    27.6 &     6.4 & 19.0 &   1.40 &   1.53 \\
 732 & J181848.5-135105 & 18 18 48.60 & -13 51 05.2 &    21 &    20.4 &     5.7 & 19.4 &   0.77 &   1.80 \\
 737 & J181848.6-134949 & 18 18 48.70 & -13 49 49.7 &     5 &     4.7 &     3.4 & \phn6.3 &   0.54 &   2.66 \\
 744 & J181848.9-134906 & 18 18 48.90 & -13 49 06.7 &    38 &    37.6 &     7.2 & 41.5 &   0.75 &   1.90 \\
 751 & J181849.2-134938 & 18 18 49.20 & -13 49 38.9 &     6 &     5.4 &     3.6 & \phn1.6 &   0.46 &   1.07 \\
 754 & J181849.3-135020 & 18 18 49.33 & -13 50 20.7 &    43 &    42.6 &     7.6 & 43.8 &   0.59 &   1.81 \\
 759 & J181849.4-134845 & 18 18 49.48 & -13 48 45.5 &   102 &   101.6 &    11.1 & 274.4\phn &   1.36 &   2.06 \\
 762 & J181849.5-135049 & 18 18 49.58 & -13 50 49.3 &    46 &    45.5 &     7.8 & 46.0 &   1.23 &   1.45 \\
 767 & J181849.7-134856 & 18 18 49.78 & -13 48 56.1 &    28 &    27.6 &     6.4 & 58.4 &   0.64 &   1.71 \\
 769 & J181849.8-135014 & 18 18 49.80 & -13 50 14.0 &    21 &    20.8 &     5.7 & 15.6 &   0.94 &   1.66 \\
 770 & J181849.8-135015 & 18 18 49.82 & -13 50 15.9 &   150 &   149.6 &    13.3 & 136.2\phn &   1.45 &   1.81 \\
 772 & J181849.8-134919 & 18 18 49.86 & -13 49 19.5 &     7 &     6.6 &     3.8 & \phn7.1 &   0.78 &   2.01 \\
 773 & J181849.8-134822 & 18 18 49.88 & -13 48 22.9 &    10 &     9.6 &     4.3 & 16.8 &   0.42 &   2.44 \\
 784 & J181850.3-134854 & 18 18 50.31 & -13 48 54.1 &   113 &   112.6 &    11.7 & 757.4\phn &   0.48 &   3.33 \\
 785 & J181850.3-134920 & 18 18 50.32 & -13 49 20.6 &    21 &    20.5 &     5.7 & 16.6 &   1.27 &   1.70 \\
 794 & J181850.6-134826 & 18 18 50.63 & -13 48 26.5 &     6 &     5.5 &     3.6 & \phn1.7 &   0.75 &   1.11 \\
 796 & J181850.6-134812 & 18 18 50.67 & -13 48 12.6 &    18 &    17.6 &     5.3 & 15.6 &   0.50 &   1.56 \\
 799 & J181850.8-134843 & 18 18 50.88 & -13 48 43.4 &    16 &    15.6 &     5.1 & 15.4 &   1.30 &   1.34 \\
 802 & J181851.1-135017 & 18 18 51.13 & -13 50 17.7 &     8 &     7.6 &     4.0 & 21.4 &   1.54 &   3.56 \\
 810 & J181851.4-135048 & 18 18 51.40 & -13 50 48.7 &    14 &    13.2 &     4.8 & 12.4 &   0.61 &   2.05 \\
 811 & J181851.4-135004 & 18 18 51.46 & -13 50 04.3 &    13 &    12.4 &     4.7 & 11.0 &   0.80 &   1.99 \\
 814 & J181851.5-134941 & 18 18 51.57 & -13 49 41.4 &    13 &    12.6 &     4.7 & 11.6 &   0.57 &   1.65 \\
 818 & J181851.7-134959 & 18 18 51.73 & -13 49 59.2 &    14 &    13.3 &     4.8 & 10.5 &   0.46 &   1.70 \\
 824 & J181851.9-134831 & 18 18 51.94 & -13 48 31.4 &     7 &     6.6 &     3.8 & \phn4.9 &   0.56 &   1.67 \\
 826 & J181852.0-134930 & 18 18 52.01 & -13 49 30.8 &    18 &    17.7 &     5.3 & 18.4 &   0.33 &   1.97 \\
 827 & J181852.1-134929 & 18 18 52.11 & -13 49 29.1 &     9 &     8.7 &     4.1 & 14.0 &   1.10 &   1.95 \\
 830 & J181852.1-135001 & 18 18 52.17 & -13 50 01.3 &    12 &    11.5 &     4.6 & \phn8.0 &   0.58 &   1.72 \\
 831 & J181852.1-135006 & 18 18 52.17 & -13 50 06.9 &    26 &    25.0 &     6.2 & 23.9 &   1.23 &   1.86 \\
 843 & J181852.6-134942 & 18 18 52.68 & -13 49 42.5 &   131 &   130.4 &    12.5 & 74.5 &   0.62 &   0.98 \\
 849 & J181852.8-134837 & 18 18 52.82 & -13 48 37.2 &     9 &     8.5 &     4.1 & 11.6 &   0.32 &   2.29 \\
 854 & J181853.0-134906 & 18 18 53.05 & -13 49 06.7 &    12 &    11.6 &     4.6 & 19.9 &   0.49 &   1.77 \\
 858 & J181853.2-134948 & 18 18 53.25 & -13 49 48.9 &    24 &    23.6 &     6.0 & 15.3 &   0.53 &   1.49 \\
 868 & J181853.8-134932 & 18 18 53.87 & -13 49 32.6 &     7 &     6.5 &     3.8 & \phn8.4 &   0.34 &   2.27 \\
 912 & J181856.2-135012 & 18 18 56.26 & -13 50 12.1 &    16 &    15.1 &     5.1 & \phn6.5 &   1.08 &   1.09 \\
\enddata
\tablecomments{Table 1 for the entire {\it Chandra} field (1103 records) is available in the electronic version of Paper~II. $f_{\rm X}$ is the 0.5-7.0 keV X-ray flux (not corrected for absorption). KS is the Kolmogorov-Smirnov time variability statistic. $\bar{e}$ is the mean photon energy.}
\end{deluxetable}

\clearpage
\pagestyle{empty}
\setlength{\voffset}{10mm}
\ptlandscape
\setlength\voffset{-1.25in}
\setlength\hoffset{-0.25in}
\begin{deluxetable}{lccccrrrcccrrrccrrcc}
\tabletypesize{\scriptsize}
\tablewidth{0pt}
\tablecolumns{18}
\small
\tablecaption{Optical and Near-Infrared Counterparts: WFPC2 Field}
\tablenum{2}
\tablehead{
\colhead{Source} &
\colhead{Designation} &
\multicolumn{1}{c}{VLT R.A.} &
\multicolumn{1}{c}{VLT Dec.} &
\colhead{Offset} &
\colhead{$J_{\rm s}$} &
\colhead{$H$} &
\colhead{$K_{\rm s}$} &
\multicolumn{1}{c}{2MASS R.A.} &
\multicolumn{1}{c}{2MASS Dec.} &
\colhead{Offset} &
\colhead{$J$} &
\colhead{$H$} &
\colhead{$K_{\rm s}$} &
\colhead{Walker} &
\colhead{Belikov} &
\colhead{$B$} &
\colhead{$V$} &
\colhead{MK} &
\colhead{$A_V$} \\
\colhead{} &
\colhead{} &
\multicolumn{2}{c}{(J2000)} &
\multicolumn{1}{c}{($\arcsec$)} &
\multicolumn{3}{c}{} &
\multicolumn{2}{c}{(J2000)} &
\multicolumn{1}{c}{($\arcsec$)} &
\multicolumn{4}{c}{} &
\multicolumn{5}{c}{} }
\startdata
 700 & J181847.4-135043 & 18 18 47.48 & -13 50 43.7 &  0.43 & 14.84 & 13.96 & 13.74 & 18 18 47.48 & -13 50 43.7 &  0.43 & 14.85 & 13.82 & 12.83 & \nodata & \nodata & \nodata & \nodata & \nodata & \nodata \\
 705 & J181847.6-134903 & \nodata & \nodata & \nodata & \nodata & \nodata & \nodata & 18 18 47.62 & -13 49 03.6 &  0.09 & 14.03 & 13.07 & 12.68 & \nodata & \nodata & \nodata & \nodata & \nodata & \nodata \\
 710 & J181847.8-135056 & 18 18 47.84 & -13 50 57.0 &  0.20 & 14.71 & 13.84 & 13.63 & 18 18 47.84 & -13 50 57.0 &  0.24 & 14.82 & 13.81 & 13.41 & \nodata & \nodata & \nodata & \nodata & \nodata & \nodata \\
 717 & J181847.9-135058 & 18 18 47.97 & -13 50 58.2 &  0.13 & 17.76 & 16.12 & 15.32 & \nodata & \nodata & \nodata & \nodata & \nodata & \nodata & \nodata & \nodata & \nodata & \nodata & \nodata & \nodata \\
 725 & J181848.2-134908 & \nodata & \nodata & \nodata & \nodata & \nodata & \nodata & 18 18 48.22 & -13 49 08.3 &  0.16 & 12.66 & 11.55 & 10.72 & \nodata & \nodata & \nodata & \nodata & \nodata & \nodata \\
 726 & J181848.2-134858 & 18 18 48.23 & -13 48 58.4 &  0.24 & 15.06 & 14.20 & 13.89 & 18 18 48.24 & -13 48 58.3 &  0.14 & 15.28 & 13.62 & 12.72 & \nodata & \nodata & \nodata & \nodata & \nodata & \nodata \\
 728 & J181848.3-134914 & 18 18 48.35 & -13 49 14.9 &  0.31 & 15.06 & 14.19 & 13.92 & 18 18 48.34 & -13 49 14.6 &  0.11 & 15.18 & 14.20 & 13.90 & \nodata & \nodata & \nodata & \nodata & \nodata & \nodata \\
 732 & J181848.5-135105 & 18 18 48.60 & -13 51 05.3 &  0.08 & 15.56 & 14.75 & 14.43 & 18 18 48.59 & -13 51 05.3 &  0.13 & 15.73 & 14.58 & 14.14 & \nodata & \nodata & \nodata & \nodata & \nodata & \nodata \\
 737 & J181848.6-134949 & \nodata & \nodata & \nodata & \nodata & \nodata & \nodata & 18 18 48.65 & -13 49 49.8 &  0.76 & 13.68 & 12.40 & 11.39 & \nodata & \nodata & \nodata & \nodata & \nodata & \nodata \\
 744 & J181848.9-134906 & 18 18 48.90 & -13 49 07.1 &  0.37 & 15.80 & 14.89 & 14.58 & 18 18 48.90 & -13 49 07.0 &  0.30 & 15.90 & 14.91 & 14.77 & \nodata & \nodata & \nodata & \nodata & \nodata & \nodata \\
 751 & J181849.2-134938 & 18 18 49.21 & -13 49 38.9 &  0.06 & 16.31 & 15.60 & 15.13 & \nodata & \nodata & \nodata & \nodata & \nodata & \nodata & \nodata & \nodata & \nodata & \nodata & \nodata & \nodata \\
 754 & J181849.3-135020 & 18 18 49.34 & -13 50 20.8 &  0.13 & 15.46 & 14.66 & 14.35 & 18 18 49.34 & -13 50 20.7 &  0.09 & 15.57 & 13.52 & 12.74 & \nodata & \nodata & \nodata & \nodata & \nodata & \nodata \\
 759 & J181849.4-134845 & \nodata & \nodata & \nodata & \nodata & \nodata & \nodata & 18 18 49.48 & -13 48 45.5 &  0.04 & 12.77 & 11.75 & 11.36 & \nodata & \nodata & \nodata & \nodata & \nodata & \nodata \\
 762 & J181849.5-135049 & 18 18 49.58 & -13 50 49.3 &  0.02 & 14.80 & 14.25 & 14.13 & 18 18 49.57 & -13 50 49.3 &  0.23 & 14.93 & 14.39 & 14.19 & \nodata & \nodata & \nodata & \nodata & \nodata & \nodata \\
 767 & J181849.7-134856 & \nodata & \nodata & \nodata & \nodata & \nodata & \nodata & 18 18 49.77 & -13 48 56.1 &  0.25 & 14.42 & 12.57 & 11.81 & \nodata & \nodata & \nodata & \nodata & \nodata & \nodata \\
 769 & J181849.8-135014 & 18 18 49.81 & -13 50 14.4 &  0.41 & 14.83 & 13.85 & 13.50 & \nodata & \nodata & \nodata & \nodata & \nodata & \nodata & \nodata & \nodata & \nodata & \nodata & \nodata & \nodata \\
 770 & J181849.8-135015 & \nodata & \nodata & \nodata & \nodata & \nodata & \nodata & 18 18 49.82 & -13 50 15.7 &  0.21 & 13.23 & 12.25 & 11.72 & \nodata & \nodata & \nodata & \nodata & \nodata & \nodata \\
 772 & J181849.8-134919 & 18 18 49.86 & -13 49 19.7 &  0.22 & 16.35 & 15.46 & 15.10 & \nodata & \nodata & \nodata & \nodata & \nodata & \nodata & \nodata & \nodata & \nodata & \nodata & \nodata & \nodata \\
 773 & J181849.8-134822 & 18 18 49.91 & -13 48 23.2 &  0.56 & 15.92 & 14.97 & 14.61 & 18 18 49.91 & -13 48 23.2 &  0.55 & 16.06 & 15.37 & 14.30 & \nodata & \nodata & \nodata & \nodata & \nodata & \nodata \\
 784 & J181850.3-134854 & 18 18 50.26 & -13 48 53.9 &  0.84 & 17.07 & 15.68 & 14.76 & \nodata & \nodata & \nodata & \nodata & \nodata & \nodata & \nodata & \nodata & \nodata & \nodata & \nodata & \nodata \\
 785 & J181850.3-134920 & 18 18 50.32 & -13 49 20.8 &  0.18 & 14.91 & 13.93 & 13.44 & \nodata & \nodata & \nodata & \nodata & \nodata & \nodata & \nodata & \nodata & \nodata & \nodata & \nodata & \nodata \\
 794 & J181850.6-134826 & 18 18 50.63 & -13 48 26.7 &  0.20 & 16.49 & 15.54 & 15.03 & \nodata & \nodata & \nodata & \nodata & \nodata & \nodata & \nodata & \nodata & \nodata & \nodata & \nodata & \nodata \\
 796 & J181850.6-134812 & \nodata & \nodata & \nodata & \nodata & \nodata & \nodata & 18 18 50.68 & -13 48 12.7 &  0.14 & \phn9.99 & \phn\nodata & \phn9.67 &   351 & 29691 & 11.72 & 11.26 & B1 & 2.65 \\
 799 & J181850.8-134843 & \nodata & \nodata & \nodata & \nodata & \nodata & \nodata & 18 18 50.87 & -13 48 43.5 &  0.17 & 13.47 & 13.16 & 12.81 &   352 & 29703 & 16.36 & 15.47 & \nodata & \nodata \\
 802 & J181851.1-135017 & 18 18 51.14 & -13 50 18.1 &  0.39 & 21.40 & 19.58 & 18.01 & \nodata & \nodata & \nodata & \nodata & \nodata & \nodata & \nodata & \nodata & \nodata & \nodata & \nodata & \nodata \\
 810 & J181851.4-135048 & 18 18 51.39 & -13 50 48.6 &  0.14 & 16.06 & 15.07 & 14.76 & \nodata & \nodata & \nodata & \nodata & \nodata & \nodata & \nodata & \nodata & \nodata & \nodata & \nodata & \nodata \\
 811 & J181851.4-135004 & 18 18 51.45 & -13 50 04.6 &  0.26 & 16.74 & 15.81 & 15.02 & \nodata & \nodata & \nodata & \nodata & \nodata & \nodata & \nodata & \nodata & \nodata & \nodata & \nodata & \nodata \\
 814 & J181851.5-134941 & 18 18 51.57 & -13 49 41.6 &  0.17 & 15.73 & 14.97 & 14.48 & \nodata & \nodata & \nodata & \nodata & \nodata & \nodata & \nodata & \nodata & \nodata & \nodata & \nodata & \nodata \\
 818 & J181851.7-134959 & 18 18 51.74 & -13 49 59.4 &  0.24 & 16.19 & 15.38 & 15.03 & 18 18 51.74 & -13 49 59.6 &  0.43 & 15.91 & 15.06 & 14.20 & \nodata & \nodata & \nodata & \nodata & \nodata & \nodata \\
 824 & J181851.9-134831 & 18 18 51.94 & -13 48 31.6 &  0.22 & 15.74 & 14.76 & 14.35 & 18 18 51.94 & -13 48 31.2 &  0.22 & 15.70 & 14.52 & 13.19 & \nodata & \nodata & \nodata & \nodata & \nodata & \nodata \\
 826 & J181852.0-134930 & 18 18 52.02 & -13 49 30.9 &  0.12 & 14.29 & 13.43 & 13.10 & \nodata & \nodata & \nodata & \nodata & \nodata & \nodata & \nodata & \nodata & \nodata & \nodata & \nodata & \nodata \\
 827 & J181852.1-134929 & \nodata & \nodata & \nodata & \nodata & \nodata & \nodata & 18 18 52.09 & -13 49 29.4 &  0.48 & 11.73 & 11.42 & 11.13 & \nodata & \nodata & \nodata & \nodata & \nodata & \nodata \\
 830 & J181852.1-135001 & 18 18 52.17 & -13 50 01.4 &  0.15 & 15.33 & 14.63 & 14.11 & 18 18 52.16 & -13 50 01.4 &  0.23 & 15.23 & 14.37 & 13.86 & \nodata & \nodata & \nodata & \nodata & \nodata & \nodata \\
 831 & J181852.1-135006 & 18 18 52.16 & -13 50 07.1 &  0.30 & 15.74 & 14.81 & 14.39 & 18 18 52.18 & -13 50 07.2 &  0.29 & 15.29 & 14.37 & 13.95 & \nodata & \nodata & \nodata & \nodata & \nodata & \nodata \\
 831 & \nodata & 18 18 52.23 & -13 50 07.4 &  0.94 & 16.15 & 15.57 & 15.39 & \nodata & \nodata & \nodata & \nodata & \nodata & \nodata & \nodata & \nodata & \nodata & \nodata & \nodata & \nodata \\
 843 & J181852.6-134942 & \nodata & \nodata & \nodata & \nodata & \nodata & \nodata & 18 18 52.67 & -13 49 42.7 &  0.23 & \phn8.80 & \phn\nodata & \phn8.76 &   367 & 29726 & \phn9.70 & \phn9.44 & O9.5 & 2.02 \\
 849 & J181852.8-134837 & 18 18 52.84 & -13 48 36.7 &  0.59 & 15.44 & 14.05 & 13.20 & 18 18 52.80 & -13 48 36.3 &  0.90 & 14.62 & 13.86 & 13.03 & \nodata & \nodata & \nodata & \nodata & \nodata & \nodata \\
 854 & J181853.0-134906 & 18 18 53.07 & -13 49 06.8 &  0.21 & 14.89 & 13.86 & 13.25 & 18 18 53.07 & -13 49 06.7 &  0.25 & 14.77 & 13.77 & 13.07 & \nodata & \nodata & \nodata & \nodata & \nodata & \nodata \\
 858 & J181853.2-134948 & 18 18 53.25 & -13 49 48.8 &  0.08 & 15.47 & 15.01 & 14.71 & 18 18 53.25 & -13 49 48.8 &  0.10 & 15.14 & 14.66 & 13.70 & \nodata & \nodata & \nodata & \nodata & \nodata & \nodata \\
 868 & J181853.8-134932 & 18 18 53.87 & -13 49 32.6 &  0.02 & 16.77 & 15.92 & 15.19 & \nodata & \nodata & \nodata & \nodata & \nodata & \nodata & \nodata & \nodata & \nodata & \nodata & \nodata & \nodata \\
 912 & J181856.2-135012 & 18 18 56.26 & -13 50 12.1 &  0.10 & 15.94 & 15.29 & 15.06 & \nodata & \nodata & \nodata & \nodata & \nodata & \nodata & \nodata & \nodata & \nodata & \nodata & \nodata & \nodata \\
\enddata
\tablecomments{Table 2 for the entire {\it Chandra} field (1161 records) is available online in machine readable format. Entries with no data in column 2 have the same designation as the source in the previous row. This X-ray source has two or more candidate near-infrared counterparts.}
\end{deluxetable}

\clearpage
\pagestyle{plaintop}
\setlength{\voffset}{0mm}
\begin{deluxetable}{lllcrrcrccccc}
\tabletypesize{\scriptsize}
\small
\tablewidth{0pt}
\tablecaption{Chandra Sources in M16: Sugitani et al. (2002) Matches}
\tablenum{3}
\tablehead{\colhead{Src} & \colhead{Designation} & \colhead{Sugitani} &
\colhead{Offset} &
\colhead{Net} & \colhead{KS} & \colhead{$\bar{e}$} &
\colhead{$f_{\rm X}$\tablenotemark{\dag}} &
\colhead{$J$} & \colhead{$H$} & \colhead{$K_{\rm s}$} &
\colhead{$A_V$} & \colhead{$\log L_{\rm X}$\tablenotemark{\ddag}} \\
\colhead{} & \colhead{} & \colhead{SNIRS} &
\colhead{($\arcsec$)} &
\colhead{(cts)} & \colhead{} & \colhead{(keV)} &
\colhead{} &
\multicolumn{4}{c}{(mag)} & \colhead{(erg s$^{-1}$)} }
\startdata
 716 & J181847.9-134836 & 181848.0-134836 &  0.21 &   36.6 &  0.49 &  1.73 &   50.9 & 13.58 & 12.50 & 11.78 &     4.1 & 30.8 \\
 726 & J181848.2-134858 & 181848.2-134858 &  0.09 &   24.6 &  0.61 &  2.04 &   38.6 & 15.23 & 14.35 & 13.97 &     4.1 & 30.5 \\
 725 & J181848.2-134908 & 181848.2-134908 &  0.25 &  119.2 &  0.87 &  1.68 &  109.9 & 12.93 & 11.80 & 11.08 &     4.5 & 31.1 \\
 728 & J181848.3-134914 & 181848.4-134915 &  0.16 &   27.6 &  1.40 &  1.53 &   19.0 & 15.20 & 14.29 & 13.93 &     4.4 & 30.5 \\
 732 & J181848.5-135105 & 181848.6-135105 &  0.16 &   20.4 &  0.77 &  1.80 &   19.4 & 15.64 & 14.71 & 14.35 &     4.5 & 30.3 \\
 737 & J181848.6-134949 &    M16ES-2~(T1) &  0.72 &    4.7 &  0.54 &  2.66 &    6.3 & 15.27 & 14.05 & 13.18 &     5.0 & 30.2 \\
 744 & J181848.9-134906 & 181848.9-134907 &  0.12 &   37.6 &  0.75 &  1.90 &   41.5 & 15.86 & 14.94 & 14.51 &     4.4 & 30.6 \\
 754 & J181849.3-135020 & 181849.4-135021 &  0.27 &   42.6 &  0.59 &  1.81 &   43.8 & 15.49 & 14.59 & 14.26 &     4.2 & 30.7 \\
 759 & J181849.4-134845 & 181849.5-134846 &  0.13 &  101.6 &  1.36 &  2.06 &  274.4 & 12.79 & 11.83 & 11.52 &     4.8 & 31.4 \\
 762 & J181849.5-135049 & 181849.6-135049 &  0.16 &   45.5 &  1.23 &  1.45 &   46.0 & 14.82 & 14.25 & 14.06 &     1.2 & 30.5 \\
 767 & J181849.7-134856 & 181849.8-134856 &  0.13 &   27.6 &  0.64 &  1.71 &   58.4 & 14.79 & 13.80 & 13.26 &     5.1 & 30.9 \\
 769 & J181849.8-135014 & 181849.8-135014 &  0.48 &   20.8 &  0.94 &  1.66 &   15.6 & 14.90 & 13.88 & 13.45 &     5.3 & 30.4 \\
 772 & J181849.8-134919 & 181849.9-134920 &  0.19 &    6.6 &  0.78 &  2.01 &    7.1 & 16.35 & 15.44 & 15.07 &     4.3 & 29.8 \\
 770 & J181849.8-135015 & 181849.8-135016 &  0.32 &  149.6 &  1.45 &  1.81 &  136.2 & 13.38 & 12.46 & 12.01 &     4.4 & 31.2 \\
 784 & J181850.3-134854 &    M16ES-1~(P1) &  0.12 &  112.6 &  0.48 &  3.33 &  757.4 & 16.84 & 15.18 & 13.59 & 27.0\phn& 32.2 \\
 785 & J181850.3-134920 & 181850.3-134921 &  0.33 &   20.5 &  1.27 &  1.70 &   16.6 & 15.09 & 14.00 & 13.45 &     6.0 & 30.5 \\
 799 & J181850.8-134843 & 181850.9-134843 &  0.09 &   15.6 &  1.30 &  1.34 &   15.4 &\nodata& 13.18 & 13.03 & \nodata & 29.9 \\
 810 & J181851.4-135048 & 181851.4-135049 &  0.18 &   13.2 &  0.61 &  2.05 &   12.4 & 16.06 & 15.11 & 14.67 &     4.7 & 30.1 \\
 811 & J181851.4-135004 & 181851.5-135004 &  0.20 &   12.4 &  0.80 &  1.99 &   11.0 & 16.75 & 15.82 & 14.97 &     2.7 & 29.9 \\
 814 & J181851.5-134941 & 181851.6-134942 &  0.21 &   12.6 &  0.57 &  1.65 &   11.6 & 15.68 & 14.77 & 14.39 &     4.3 & 30.2 \\
 815 & J181851.6-135127 & 181851.6-135128 &  0.20 &  125.0 &  2.58 &  1.86 &  121.8 & 16.08 & 15.20 & 14.84 &     4.1 & 31.1 \\
 818 & J181851.7-134959 & 181851.8-134959 &  0.31 &   13.3 &  0.46 &  1.70 &   10.5 & 16.27 & 15.33 & 14.95 &     4.5 & 30.1 \\
 826 & J181852.0-134930 & 181852.0-134931 &  0.38 &   17.7 &  0.33 &  1.97 &   18.4 & 14.23 & 13.35 & 12.92 &     4.1 & 30.2 \\
 827 & J181852.1-134929 & 181852.1-134929 &  0.12 &    8.7 &  1.10 &  1.95 &   14.0 & 11.94 & 11.55 & 11.44 & \nodata & 29.8 \\
 830 & J181852.1-135001 & 181852.2-135001 &  0.12 &   11.5 &  0.58 &  1.72 &    8.0 & 15.34 & 14.50 & 13.95 &     1.8 & 29.8 \\
 831 & J181852.1-135006 & 181852.2-135007 &  0.21 &   25.0 &  1.23 &  1.86 &   23.9 & 16.25 & 15.52 & 15.31 &     2.6 & 30.3 \\
 843 & J181852.6-134942 & BD~-13~4930\tablenotemark{\ast} & 0.14 &  130.4 &  0.62 &  0.98 &   74.5 &\phn8.79 &\phn8.81 &\phn8.77 &     1.0 & 30.8 \\
 849 & J181852.8-134837 & 181852.9-134837 &  0.78 &    8.5 &  0.32 &  2.29 &   11.6 & 15.40 & 14.30 & 13.50 &     4.2 & 30.0 \\
 854 & J181853.0-134906 & 181853.1-134907 &  0.38 &   11.6 &  0.49 &  1.77 &   19.9 & 14.81 & 13.80 & 13.23 &     5.3 & 30.4 \\
 858 & J181853.2-134948 & 181853.3-134949 &  0.19 &   23.6 &  0.53 &  1.49 &   15.3 & 15.36 & 14.80 & 14.62 &     1.0 & 30.0 \\
 868 & J181853.8-134932 & 181853.9-134933 &  0.10 &    6.5 &  0.34 &  2.27 &    8.4 & 16.70 & 15.70 & 15.13 &     5.1 & 29.9 \\
 866 & J181853.8-135127 & 181853.9-135128 &  0.20 &   57.0 &  4.42 &  2.29 &   83.0 & 15.64 & 14.65 & 14.19 &     5.1 & 30.9 \\
 892 & J181855.3-135049 & 181855.4-135050 &  0.89 &   11.5 &  1.35 &  1.75 &    9.2 & 16.69 & 15.25 & 14.53 &     9.3 & 30.6 \\
 898 & J181855.5-134844 & 181855.6-134844 &  0.10 &  100.5 &  0.91 &  1.75 &  103.6 & 11.23 & 10.41 & 10.56 &     3.4 & 31.0 \\
 903 & J181855.9-135048 & 181855.9-135048 &  0.26 &   63.0 &  2.90 &  2.19 &   77.0 & 15.00 & 14.00 & 13.57 &     5.1 & 30.9 \\
 912 & J181856.2-135012 & 181856.3-135012 &  0.12 &   15.1 &  1.08 &  1.09 &    6.5 & 15.89 & 15.30 & 14.98 &     1.3 & 29.8 \\
 919 & J181856.6-134937 & 181856.7-134938 &  0.31 &   53.4 &  0.97 &  1.45 &   34.1 & 15.03 & 14.16 & 13.83 &     3.9 & 30.7 \\
 929 & J181857.3-134842 & 181857.4-134842 &  0.32 &    8.3 &  0.55 &  4.74 &   46.0 & 16.73 & 15.87 & 15.39 &     3.9 & 30.1 \\
 939 & J181857.9-134905 & 181858.0-134906 &  0.47 &   11.9 &  0.89 &  1.77 &   15.9 & 15.03 & 14.16 & 13.81 &     3.9 & 30.2 \\
 941 & J181857.9-135123 & 181858.0-135124 &  0.33 &    8.7 &  0.48 &  1.34 &    4.9 & 15.16 & 14.30 & 13.93 &     3.8 & 30.0 \\
\enddata
\tablenotetext{\dag}{$f_{\rm X}$ is the absorbed 0.5-7 keV flux at Earth in units of $10^{-16}$ ergs cm$^{-2}$ s$^{-1}$.}
\tablenotetext{\ddag}{$L_{\rm X}$ is the unabsorbed 0.5-7 keV luminosity assuming $d=2$~kpc.}
\tablenotetext{\ast}{For the bright star BD~-13~4930 photometry is from 2MASS, astrometry is from Hipparcos.}
\end{deluxetable}

\clearpage
\begin{deluxetable}{lcccclcccc}
\tabletypesize{\scriptsize}
\tablewidth{0pt}
\tablecolumns{10}
\small
\tablecaption{Comparison of Stellar X-ray Properties}
\tablenum{4}
\tablehead{
\colhead{Star} &
\colhead{Spectral} &
\colhead{Age} &
\colhead{Star Formation} &
\colhead{Magnetic} &
\colhead{$T$} &
\colhead{$\log L_{\rm X}^{\rm total}$} &
\colhead{$\log L_{\rm X}^{\rm wind}$} &
\colhead{$\log L_{\rm X}^{\rm total}/L_{\rm bol}$} &
\colhead{References\tablenotemark{b}} \\
\colhead{} &
\colhead{Type} &
\colhead{Myr} &
\colhead{Region} &
\colhead{Field} &
\colhead{(keV)} &
\colhead{} &
\colhead{} &
\colhead{} &
\colhead{}}
\startdata
M16ES-1 & O & PMS  & M16 & likely & $2.2^{+1.0}_{-0.6}$ & 32.2 & 
  (28.7)\tablenotemark{a} & --3.7 & 1\\
$\theta^1$ Ori C & O6.5Vp & 0.3 & Orion & yes & 0.5--5.3 & 32.3 &
  31.5 & --6.4 & 2\\
Oph S1 & B4+k &   & $\rho$ Oph & yes & $2.38\pm 0.22$ & 30.4 & 
  (29.4)\tablenotemark{a} & --4.2 & 3,4\\
$\zeta$ Ori & O9.5Ib & $<12$ & Orion & weak? & 0.16--0.8 & 32.5 &
  32.5 & --6.8 & 5\\
$\tau$ CMa & O9Ib & 3--5 & NGC 2362 & no & 0.24--0.97 & 32.2 &
  32.2 & --7.2 & 2\\
\enddata

\tablenotetext{a}{$L_{\rm X}^{\rm wind}$ was computed assuming that 
$\log L_{\rm X}^{\rm wind}/L_{\rm bol} = -7.2$.}

\tablenotetext{b}{References: (1) This paper. (2) \citet{Schulz2003}.
(3) \citet{Andre1988}. (4) \citet{Gagne2004}. (5) \citet{Waldron2000}.}

\end{deluxetable}

\clearpage
\begin{deluxetable}{lccccccc}
\tabletypesize{\scriptsize}
\tablewidth{0pt}
\tablecolumns{8}
\small
\tablecaption{Comparison of M16 EGGs to ONC YSOs} 
\tablenum{5}
\tablehead{
\colhead{EGG} &
\colhead{Mass} &
\colhead{$A_{\rm V}$} &
\colhead{$N_{\rm H}$} &
\colhead{$f_{\rm X}$} &
\colhead{$f_{\rm X}^{\rm corr}$} &
\colhead{$\log L_{\rm X}$} &
\colhead{frac\tablenotemark{a}} \\
\colhead{Number} &
\colhead{($M_{\odot}$)} &
\colhead{(mag)} &
\colhead{($10^{22}$ cm$^{-2}$)} &
\colhead{($10^{-16}$ cgs)} &
\colhead{($10^{-16}$ cgs)} &
\colhead{} &
\colhead{}}
\startdata
E25 & 0.50 & ~9 & 1.4 & 7.6 & 19.9 & 30.00 & 0.68\\
E31 & 0.35 & 10 & 1.6 & 7.8 & 29.6 & 30.15 & 0.30\\
E35 & 0.95 & 22 & 3.5 & 8.1 & 52.4 & 30.40 & 0.41\\
E42 & 1.00 & ~4 & 0.6 & 5.9 & 13.4 & 29.80 & 0.89\\
    &      &    &     &     &      &       &     \\
    &      &    &     &     &      & Sum = & 2.28\\
\enddata

\tablenotetext{a}{The approximate fraction of YSOs in the appropriate 
mass range with $L_{\rm X}$ above this limit in the COUP survey.}

\end{deluxetable}

\clearpage

\begin{figure}
\plotone{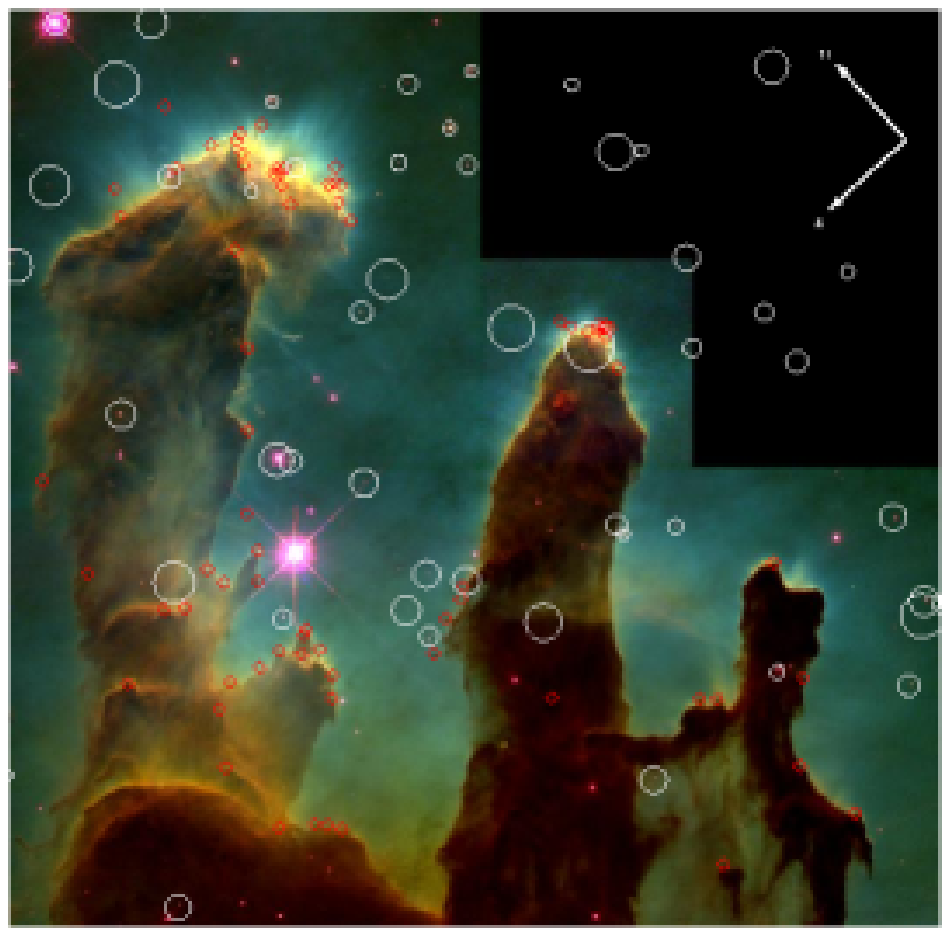}
\caption{
1995 HST WFPC2 image of the $2.5\arcmin\times2.5\arcmin$ region centered on 
J2000
$(\alpha,\delta) = 18^{\mathrm h}18^{\mathrm m}50\fs6, 
-13^\circ49\arcmin52\arcsec$,
including (from left to right) Pillars 1, 2, and 3 \citep{Hester1996}.
North is to the upper-left and east is to the lower-left.
The colors represent $H\alpha$ (red), [\ion{S}{2}] + continuum (green), 
and [\ion{O}{3}] 
(blue) emission.
The $3\sigma$ X-ray position uncertainty circles (black in the black and white 
figure and white in the color figure) indicate 
{\it Chandra} detections, the diamonds (white in the black and white figure 
and red in the color figure) indicate 
the locations of the 73 EGGs identified by \citet{Hester1996}. 
}
\end{figure}

\begin{figure}
\plotone{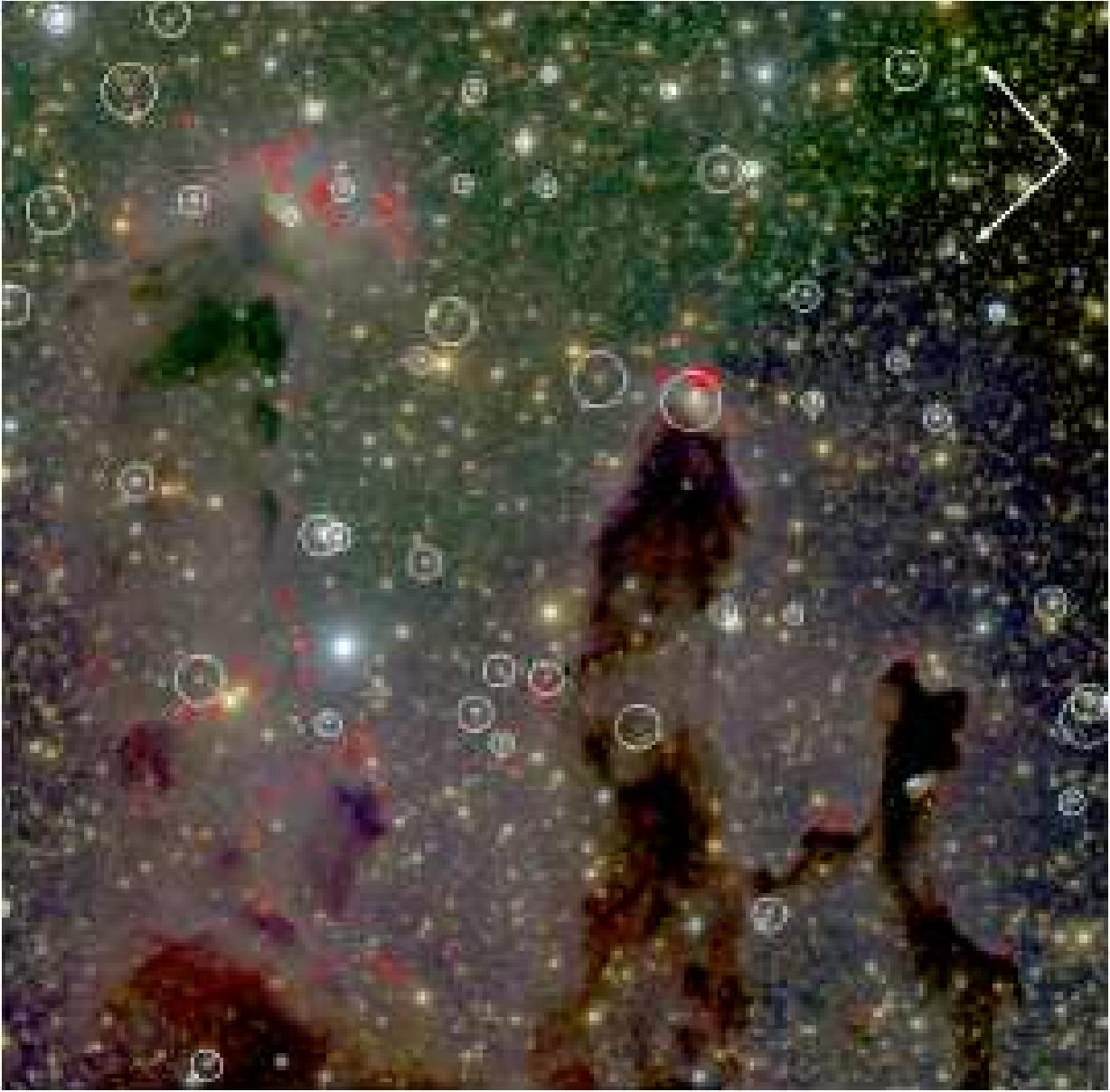}
\caption{2001 VLT ISAAC $K_{\rm s}$, $H$, $J_{\rm s}$ RGB image \citep{MA2002} 
of the same field as Figure 1 with $3\sigma$ X-ray position uncertainty 
circles and locations of EGGs (diamonds) superimposed. Of the 40 {\it Chandra}
sources in the WFPC2 field (Table~1), none are coincident with the EGGs. 
All but one {\it Chandra} source is identified with a single 2MASS or VLT 
near-infrared counterpart (see Table~2). M16ES-1 is the strong X-ray source 
(small error circle) at the head of Pillar 1 and 
M16ES-2 is the weak X-ray source (large error circle) at the head of Pillar 2.
Both are yellow in the color figure. 
}
\end{figure}

\begin{figure}
\plotone{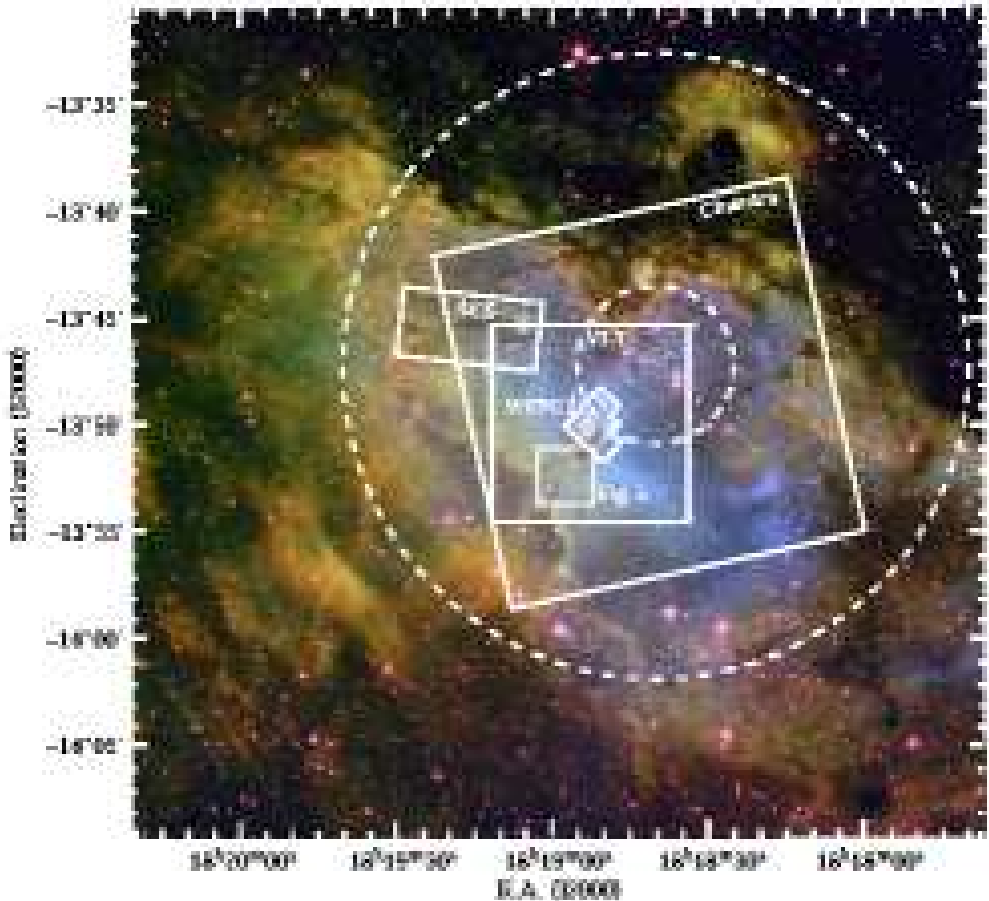}
\caption{An optical image of the NGC~6611 region showing the 
emission nebula, the optically bright OBA stars, and the molecular clouds 
surrounding the cluster core. 
The inner and outer dashed circles are the $3\sigma$ core and corona of 
NGC~6611 \citep{Belikov1999}.  
The O stars responsible for the photoevaporation
of the pillars are located near the center of NGC~6611.
Superimposed on the optical image are the boundaries of the
$16.9\arcmin\times16.9\arcmin$ 
{\it Chandra} ACIS-I field, the
$9.1\arcmin\times9.1\arcmin$ VLT ISAAC mosaic of Pillars 1--4, the
HST WFPC2 image of Pillars 1--3, the HST ACS mosaic of Pillar 5,
and the boundary of Figure 5 containing Pillar 4.
}
\end{figure}

\begin{figure}
\plotone{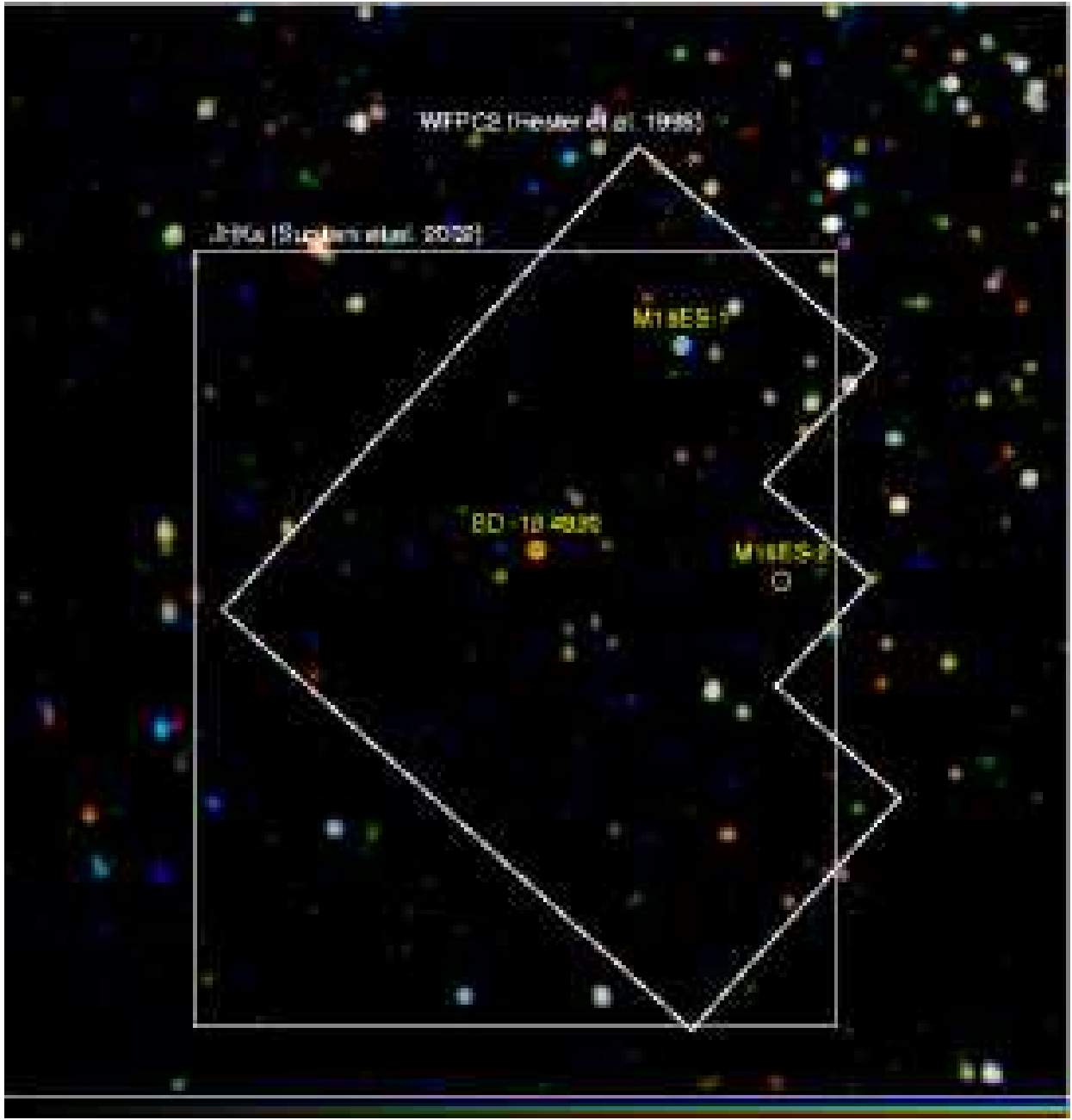}
\caption{
A portion of the 76-ks {\it Chandra} ACIS-I X-ray image of M~16 
Pillars 1, 2, and 3
obtained in 2001 June. Red, green and blue colors represent counts
in the 0.5-1.5, 1.5-2.5 and 2.5-7 keV bands.
This $4\arcmin\times4\arcmin$ region is centered on the O9.5 star
BD~-13 4930 at J2000 $(\alpha,\delta) =
18^{\rm h}18^{\rm m}52\fs7, -13^\circ49\arcmin43\arcsec$.
North is up and east is to the left. Included are the
outlines of the \citet{Sugitani2002} $JHK_{\rm s}$ field and
the \citet{Hester1996} {\it HST} WFPC2 field. Prominent stars 
in Table~3 are circled and labeled in yellow.
The hot stars of NGC~6611 are located in the northwest corner of
the image. Soft X-ray sources like BD~-13 4930 appear red and hard,
deeply embedded YSOs appear blue.
The full-resolution ($0.492\arcsec$) ACIS grayscale image (RGB scale in the 
color image) was smoothed with a Tophat function with 
kernal radius of three pixels. 
}
\end{figure}

\begin{figure}
\plotone{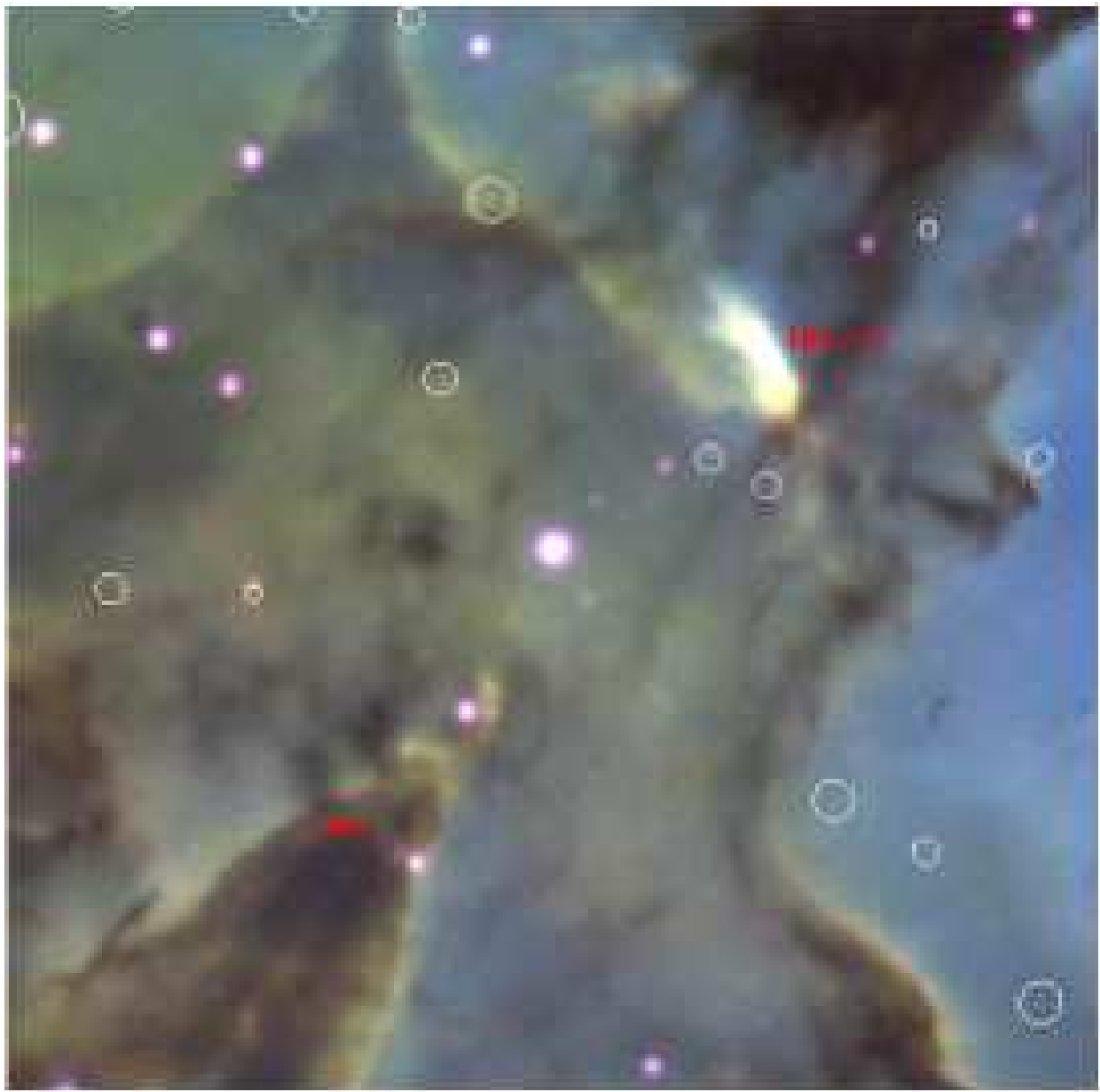}
\caption{
NOAO 0.9-m MOSA [\ion{S}{2}] (red), $H\alpha$ (green), and [\ion{O}{3}] (blue) 
image of the $2.5\arcmin\times2.5\arcmin$ region centered on J2000 
$(\alpha,\delta) = 18^{\mathrm h}18^{\mathrm m}57\fs4, 
-13^\circ52\arcmin11\arcsec$,
showing Pillar 4 and HH216. The boundaries of this field are shown in Figure~3.
The $3\sigma$ {\it Chandra} position uncertainty circles indicate the location
of X-ray sources. The location of the jet 
region in the infrared image (not visible in this image) is shown.
}
\end{figure}

\begin{figure}
\plotone{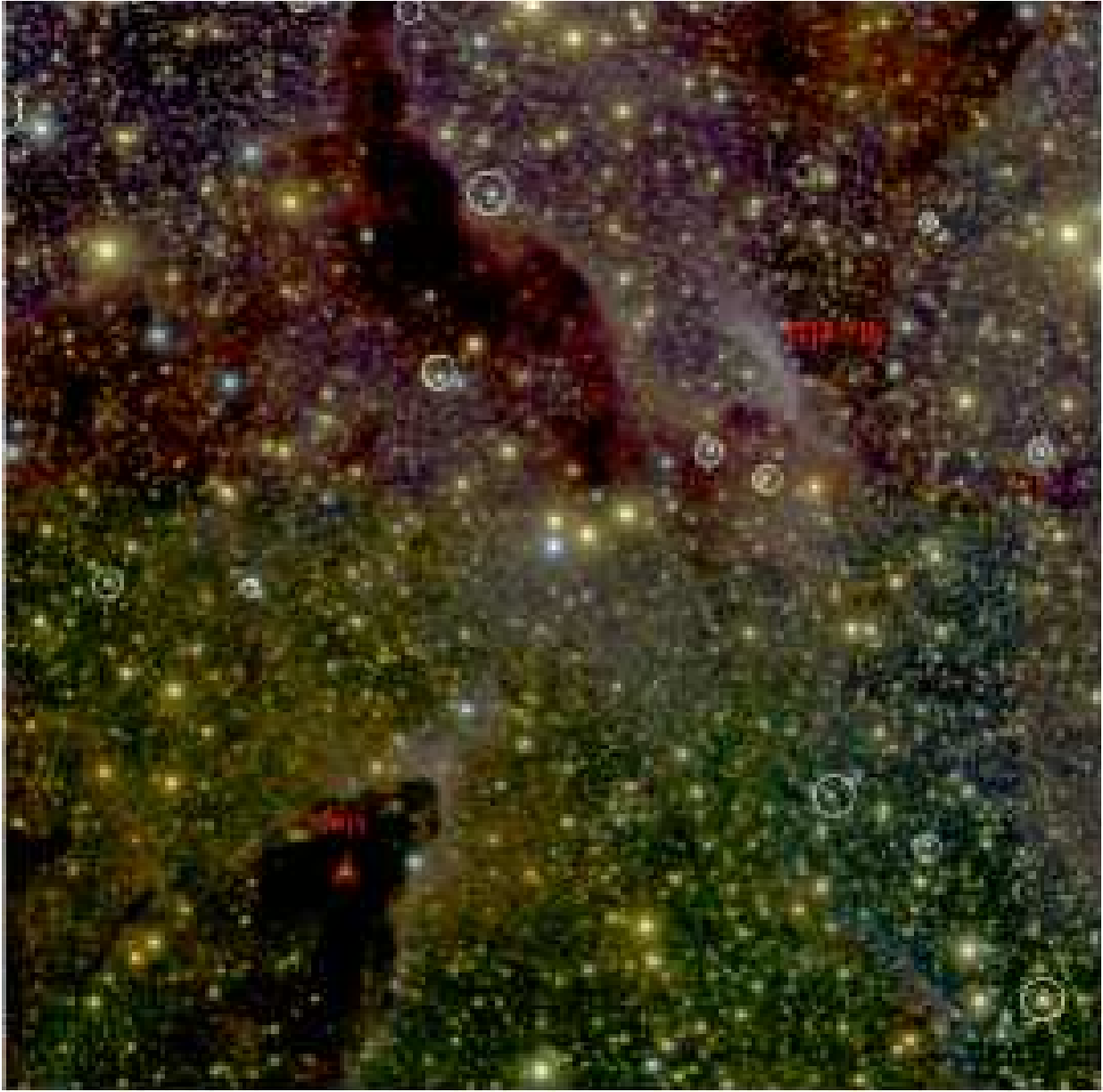}
\caption{
2001 VLT ISAAC $K_{\rm s}HJ_{\rm s}$ RGB image \citep{MA2002}
of the same field as the previous figure.
The white circles are {\it Chandra} source circles. All but one
{\it Chandra} source is identified with a single 2MASS or VLT near-infrared
countepart (see Table 2).
}
\end{figure}

\begin{figure}
\plotone{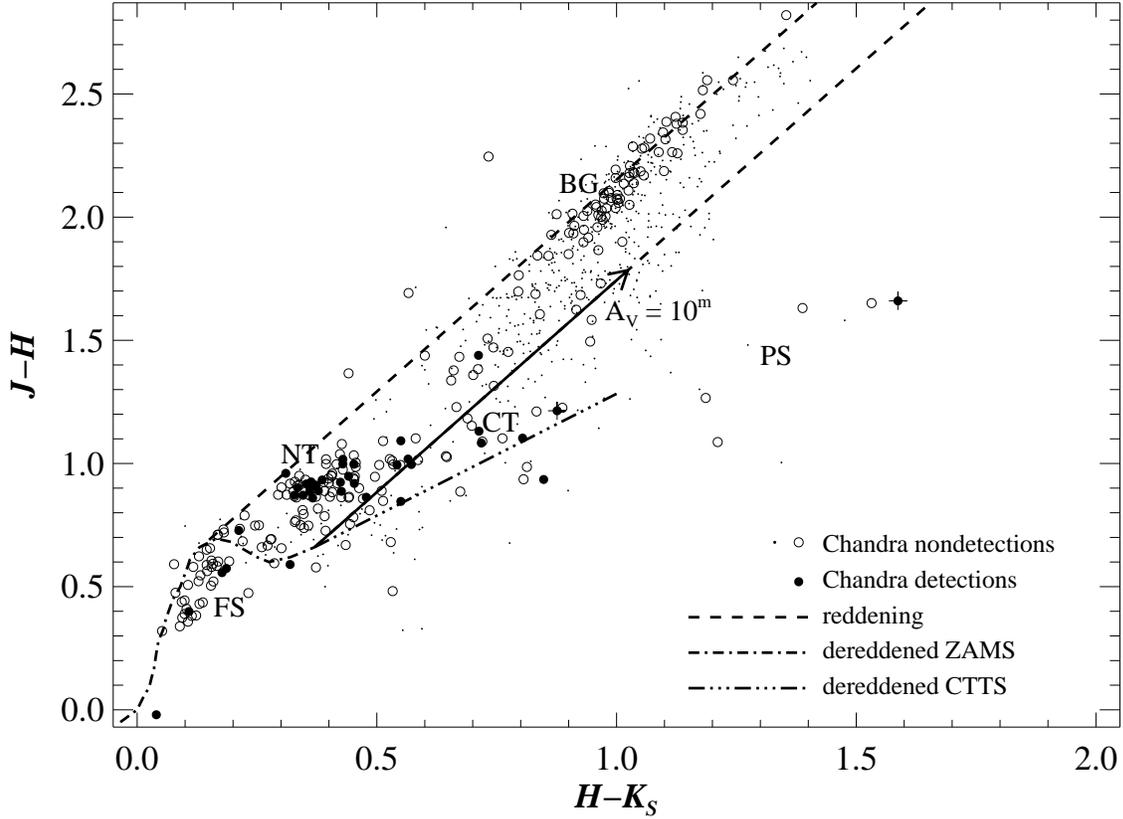}
\caption{
$JHK_{\rm s}$ color-color diagram for 784 stars in the \citet{Sugitani2002} 
field
including Pillars 1, 2 and 3. The 40 {\it Chandra}-detected stars in Table~3
are shown as large filled symbols. Large circles represent bright stars with
$K_{\rm s} \leq 17$; dots represent fainter stars. Also shown is the locus
of dereddened ZAMS stars (dash-dot line), the locus of dereddened 
classical T Tauri stars (dash-triple dot line), reddening lines
for cool giants and dwarfs (dashed lines), and an $A_V=10$~mag
extinction vector (solid arrow).
The \citet{Sugitani2002} region contains four populations of stars.
1) At $J-H \lesssim 0.8$ we see a small population
of relatively bright, unreddened ($A_V < 1$)
foreground stars (labelled FS), 
most of which are not detected with {\it Chandra}.
2) At $J-H \lesssim 1.2$ and between the dashed lines we see a population
of relatively bright, lightly reddened ($A_V\approx 2.5$) stars,
approximately $\slantfrac{1}{3}$ of which are detected with {\it Chandra}.
These are probably naked T Tauri stars (labelled NT).
3) To the left of the dashed redenning lines, above
the dereddened CTTS line, is a small population of CTTSs (labelled CT,
the ``T'' sources of \citet{Sugitani2002}). Approximately
$\slantfrac{1}{3}$ of these are detected with {\it Chandra}.
(4) At $H-K_{\rm s} \gtrsim 1$ is a small group of reddened
stars with large infrared excess (labelled PS, including Sugitani et al.'s 
protostellar ``P'' sources). Of these 8 or so stars,
only P1=M16ES-1 (large circle with a +) is detected with {\it Chandra}.
(5) A large population of highly reddened ($A_V\gtrsim 6$) stars
lies close to the upper reddening line, none of which are
detected with {\it Chandra}. These are background, galactic giants 
(labelled BG).}
\end{figure}

\end{document}